\documentclass[12pt]{article}

\newcommand{\blind}{0}

\usepackage[T1]{fontenc}
\usepackage{microtype}
\usepackage{amsfonts,amssymb,amsbsy,amsmath,amsthm}
\usepackage{mathtools}
\usepackage[onehalfspacing]{setspace}
\usepackage{booktabs}
\usepackage[inline]{enumitem}
\usepackage{graphicx,tikz,float}
\usepackage{longtable,relsize}
\usepackage[section]{placeins}

\usepackage{natbib}
\setcitestyle{aysep={}}

\usepackage{authblk}
\usepackage{hyperref}
\usepackage[margin={1in,1in},includefoot]{geometry}
\usepackage[capitalize,sort]{cleveref}

\usepackage[normalem]{ulem}

\usepackage{array}
\newcolumntype{H}{>{\setbox0=\hbox\bgroup}c<{\egroup}@{}}

\usetikzlibrary{external,arrows,positioning,decorations.pathreplacing}
\hypersetup{breaklinks,colorlinks,allcolors=blue,pdfpagemode=UseNone}

\newtheorem{proposition}{Proposition}[section]

\setlist[itemize]{leftmargin=*}

\newlist{assume}{enumerate}{1}
\setlist[assume]{label=\textup{A\arabic*},leftmargin=*,resume=assume}
\crefname{assumei}{}{}

\newlist{assumer}{enumerate}{1}
\setlist[assumer]{label=\textup{R\arabic*},leftmargin=*,resume=assumer}
\crefname{assumeri}{}{}

\newlist{assumec}{enumerate}{1}
\setlist[assumec]{label=\textup{C\arabic*},leftmargin=*,resume=assumec}
\crefname{assumeci}{}{}

\newlist{assumew}{enumerate}{1}
\setlist[assumew]{label=\textup{W\arabic*},leftmargin=*,resume=assumew}
\crefname{assumewi}{}{}

\Crefname{appendix}{Supplement}{Supplements}
\Crefname{page}{page}{pages}
\crefname{equation}{}{}
\crefname{enumi}{}{}

\makeatletter
\DeclareRobustCommand\citepos
  {\begingroup
   \def\NAT@nmfmt##1{{\NAT@up##1's}}%
   \NAT@swafalse\let\NAT@ctype\z@\NAT@partrue
   \@ifstar{\NAT@fulltrue\NAT@citetp}{\NAT@fullfalse\NAT@citetp}}
\g@addto@macro\@floatboxreset{\small\centering}
\newcommand{\tr}{{\mathpalette\@tr{}}}
\newcommand{\@tr}[2]{\raisebox{\depth}{$\m@th#1\mathsmaller\intercal$}}
\makeatother

\allowdisplaybreaks
\DeclareMathOperator{\E}{E}
\DeclareMathOperator{\pr}{Pr}

\DeclareMathOperator{\cor}{Corr}
\DeclareMathOperator{\expit}{expit}

\DeclareMathOperator{\indic}{1}
\DeclareMathOperator{\norm}{N}
\newcommand{\bigmid}{\mathop{\bigl|}}
\newcommand{\Bigmid}{\mathop{\Bigl|}}

\newcommand{\eprob}{\mathop{\mathbb{P}_n}}
\newcommand{\prob}{\mathop{\kern0pt P}}

\let\originalleft\left
\let\originalright\right
\renewcommand{\left}{\mathopen{}\mathclose\bgroup\originalleft}
\renewcommand{\right}{\aftergroup\egroup\originalright}
\mathchardef\breakingcomma\mathcode`\,{%
  \catcode`,=\active\gdef,{\breakingcomma\discretionary{}{}{}}}
\newcommand{\mathlist}[1]{\mathcode`\,=\string"8000 #1}
\newcommand{\bftab}{\fontseries{b}\selectfont}

\title{Assessing Time-Varying Causal Effect Moderation in Mobile Health}
\date{}

\author[1]{Audrey Boruvka}
\author[2]{Daniel Almirall}
\author[3]{Katie Witkiewitz}
\author[1,2]{Susan A. Murphy}

\affil[1]{Department of Statistics, University of Michigan}
\affil[2]{Institute for Social Research, University of Michigan}
\affil[3]{Department of Psychology, University of New Mexico}

\if1\blind
\hypersetup{draft}
\makeatletter\def\@author{}\makeatother
\fi

\begin{document}
\maketitle

\begin{abstract}
In mobile health interventions aimed at behavior change and maintenance, treatments are provided in real time to manage current or impending high risk situations or promote healthy behaviors in near real time. Currently there is great scientific interest in developing data analysis approaches to guide the development of mobile interventions.  In particular data from mobile health studies might be used to examine effect moderators---individual characteristics, time-varying context or past treatment response that moderate
 the effect of current treatment on a subsequent response.
This paper introduces a formal definition for moderated  effects in terms of potential outcomes, a definition  that is particularly suited to mobile interventions, where treatment occasions are numerous, individuals are not always available for treatment, and potential moderators might be influenced by past treatment.  Methods for estimating moderated effects are developed and compared.    The proposed approach is illustrated using BASICS-Mobile, a smartphone-based intervention designed to curb heavy drinking and smoking among college students.

\end{abstract}

\newpage
\section{Introduction}
\label{sec:intro}

Mobile health (mHealth) broadly refers to the practice of healthcare using mobile devices, such as smartphones and wearable sensors both to deliver treatment as well as to sense the current context  of the individual.
 In mobile interventions for behavior maintenance or change, treatments are typically designed to help individuals manage high risk situations or promote healthy behaviors.  Examples include medication reminders, motivational messages, physical activity suggestions, cognitive exercises to help manage stress or other risky situations, and prompts to  facilitate activity in support networks.

There is intense interest in data analysis approaches to guide the development of mobile interventions \citep{free2013,muessig2013} and to test the dynamic behavioral theories on which these interventions are based \citep{spring2013,mohr2014}.  Micro-randomized trials (MRTs; \citealp{klasjna2015, liao2015, dempsey2015}) provide data expressly for this purpose, with each participant in an MRT sequentially randomized to treatment numerous times, at possibly 100s to 1000s of occasions.  In both MRTs and observational mHealth studies both treatment and measurement occur intensively over time.   Measurements on individual characteristics, context and response to treatments are collected passively through sensors or actively by self-report.

One way in which these data may aid the design of a mobile intervention is through the examination of effect moderation; that is, inference about which factors strengthen or weaken the response to treatments.  Consider, for example, an intervention for smoking cessation.  Mindfulness-based treatments to help individuals manage their urge to smoke are presumably best delivered at times when there exists an inclination to smoke \citep[e.g.][]{witkiewitz2014}.  However other factors might influence the effect of these treatments on subsequent smoking rate.  For example it may be that the mindfulness-based approach reduces smoking only when stress levels or self-regulatory demands are low, and has little to no effect otherwise.  In general knowledge about moderators can be used to deliver treatments only in settings where they have proven most efficacious or to identify alternative treatment strategies when the treatment shows little to no benefit.  Treatment effects might also evolve over the course of the intervention, so functions of time could also be examined as possible moderators.

This paper provides two main contributions in the assessment of treatment effects from longitudinal data in which treatment, response, and potential moderators are time-varying.  The first is a definition for treatment effects that is particularly suited for mHealth, where treatment occasions are numerous and potential moderators might be influenced by past treatment.  These effects are a marginal generalization of the  treatment ``blips'' in the structural nested mean model (SNMM; \citealp{robins1989,robins1994,robins1997a}); the effects are conditional on a few select variables representing  potential moderators of interest as opposed to requiring that the effects be conditional on all past observed variables.   The second contribution is a centered and weighted least squares method for estimating these treatment effects. 

The most common estimation methods used in the analysis of mobile health data are generalized estimating equation (GEE) approaches or related approaches that employ random effects \citep{schafer2006, Stone2007, bolger2013}; these methods are frequently used to better understand the time-varying relationship between two variables such as craving and stress. Unfortunately, when the mobile health data includes time-varying treatment, these methods are not guaranteed to consistently estimate causal treatment effects. In this paper, we provide a centered and weighted least squares estimation method that provides unbiased estimation. 

We begin by defining treatment effects in our setting.  The centered and weighted estimation method is derived and its properties are assessed numerically using a variety of simulation scenarios. As an illustration, we apply the proposed method to data from a study of BASICS-Mobile, a mobile intervention to curb heavy drinking and smoking among college students \citep{witkiewitz2014}.

\section{Proximal and Other Lagged Treatment Effects}

\subsection{Motivating Example}
\label{sec:basics}

Our motivating example is drawn from BASICS-Mobile, a smartphone-based intervention designed to reduce heavy drinking and smoking among college students.  Users are prompted three times per day (morning, afternoon and evening) to complete a self-report assessing a variety of individual and contextual factors including episodes of drinking or smoking, social settings, affect, and need to self-regulate thoughts.  The afternoon and evening self-reports are possibly followed by a treatment module of three to four screens of information and at least one question to confirm that the module was received.  Some of the treatment modules address smoking and heavy drinking using mindfulness messages (\citealp{bowen2009}).  Other modules provide general (primarily health-related) information (\citealp{dimeff1999}).  In an analysis of data arising from the implementation of BASICS-Mobile, it is natural to estimate the effect of providing the mindfulness messages (versus providing general health information) on a proximal response, such as the smoking rate between the current and following self-report, and to assess whether or not these effects differ according to the individual's context.

\subsection{Notation and Data}
\label{sec:notation}

 For a given individual, let $A_t$ denote the treatment at the $t$th treatment occasion and $Y_{t+1}$ be the subsequent proximal response ($t = 1, \ldots, T$).  Throughout we limit attention to the case where each $A_t$ is binary and $Y_{t+1}$ is continuous.  Individual and contextual information at the $t$th treatment occasion is represented by $X_t$, which may contain summaries of previous measurements of context, treatment or response.  For example, prior to each treatment occasion the individual might report their current mood.  The vector $X_t$ could then contain this measurement or, with previous measurements, variation or change in mood.  Over the course of $T$ treatment occasions, the resulting data from an individual ordered in time is $\mathlist(X_1, A_1, Y_2, \ldots, X_T, A_T, Y_{T+1})$.  The overbar is used to denote a sequence of random variables or realized values through a specific treatment occasion; for example $\bar A_t = (A_1, \ldots, A_t)$.  Information accrued up to treatment occasion $t$ is represented by the history $H_t = (\bar X_t, \bar Y_t, \bar A_{t-1})$.

\begin{figure}
\begin{tikzpicture}[anchor=base,align=center]
\draw[-|,semithick] ( 0.00, 0) -- ( 2.00, 0)
  node[at start, left=3pt] {$\ldots$}
  node[at end, below=20pt] {Morning};
\draw[ -|,semithick] ( 2.00, 0) -- ( 5.00, 0)
  node[at end, above=6pt] {$A_{t-1}$}
  node[at end, below=6pt] {$t-1$}
  node[at end, below=20pt] {Afternoon};
\draw[ -|,semithick] ( 5.00, 0) -- ( 8.00, 0)
  node[at end, above=6pt] {$A_t$}
  node[at end, below=6pt] {$t$}
  node[at end, below=20pt] {Evening};
\draw[ -|,semithick] ( 8.00, 0) -- (11.00, 0)
  node[at end, below=20pt] {Morning};
\draw[ -,semithick] (11.00, 0) -- (13.00, 0)
  node[at end, right=3pt] {$\ldots$};
\draw[decorate,decoration={brace,amplitude=8pt,raise=24pt}] (2.05,0) -- (4.95,0)
  node[midway,above=34pt] {$X_{t-1}$};
\draw[decorate,decoration={brace,amplitude=8pt,raise=24pt}] (5.05,0) -- (7.95,0)
  node[midway,above=34pt] {$Y_t, X_t$};
\draw[decorate,decoration={brace,amplitude=8pt,raise=24pt}] (8.05,0) -- (10.95,0)
  node[midway,above=34pt] {$Y_{t+1}$};
\end{tikzpicture}
\caption{A BASICS-Mobile participant's data for two treatment occasions leading up to $Y_{t+1}$, depicted in chronological order.  Information is primarily collected via self-reports three times per day---morning, afternoon and evening.  Treatment occasions take place after the afternoon and evening self-reports.}
\label{fig:timeline}
\end{figure}
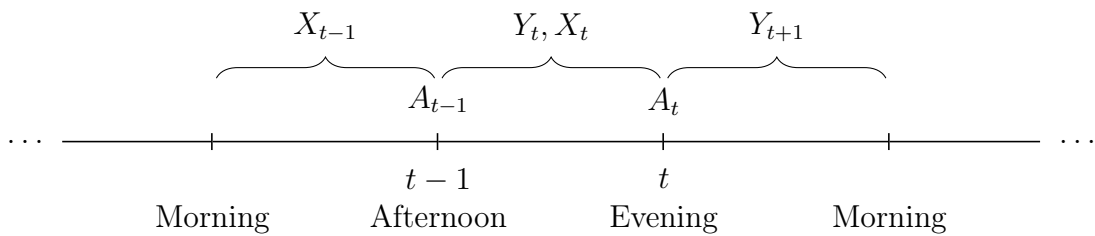

In BASICS-Mobile (\cref{fig:timeline}), $A_t = 1$ if a mindfulness message is provided at the $t$th treatment occasion and $A_t = 0$ otherwise, $Y_{t+1}$ is the smoking rate between the occasion $t$ self-report prompt and the following self-report prompt, $T=28$, and $X_t$ includes the time of day, number of reports recently completed, prior smoking rate, current need to self-regulate, and other summary variables formed from the reports up to and including the $t$th occasion.  For example, from the self-reports at $t-1$ and $t$, we can examine the change in self-regulation needs and determine whether there was an increased need ($\textit{incr}_t = 1$) or not ($\textit{incr}_t = 0$).

To define treatment effects below, we adopt potential outcomes \citep{rubin1974, neyman1990, robins1989} notation.  However we will deviate slightly from this framework because, as will be seen below in \cref{eq:lagk},  our estimands may  involve the treatment distribution in the data. We represent random variables or vectors with uppercase letters; lowercase letters denote their realized values.
 In particular it will be useful to include in the set of potential outcomes, treatments expressed as potential outcomes of past treatment.  That is, the potential outcomes are  $\mathlist{\{Y_2(a_1), X_2(a_1), A_2(a_1)\}_{a_1\in\{0,1\}}, \ldots, \{Y_T(\bar a_{T-1}), X_T(\bar a_{T-1}), A_T(\bar a_{T-1})\}_{\bar a_{T-1}\in\{0,1\}^{T-1}}, \{Y_{T+1}(\bar a_{T})\}_{\bar a_{T}\in\{0,1\}^{T}}}$.  In BASICS-Mobile, for example, the smoking rate measured following the second treatment occasion has four potential outcomes: $Y_3(0, 0)$, $Y_3(0, 1)$, $Y_3(1, 0)$, $Y_3(1, 1)$.  Here $Y_3(0, 0)$ is the smoking rate that would arise for a given individual had that individual received no mindfulness treatments over the first two treatment occasions: $a_1 = a_2 = 0$.  This idea can be similarly applied to the measurements $X_t$, since they might also be influenced by past treatment; $X_{t+1}(\bar a_t)$ are the potential measurements had the sequence of treatments $\bar a_t$ been allocated.  For brevity, we denote $A_2(A_1)$ by $A_2$ and so on with $A_t(\bar A_{t-1})$ denoted by $A_t$.  Then  $H_t(\bar A_{t-1}) = \mathlist{(X_1, A_1, Y_2(A_1), X_2(A_1), A_2, Y_3(\bar A_2), X_3(\bar A_2), A_3, \ldots, Y_t(\bar A_{t-1}), X_t(\bar A_{t-1}))}$.

\subsection{Moderated Treatment Effects}
\label{sec:effects}

Many treatments are designed to influence an individual in the short term or proximally in time \citep{heron2010}.  For example, instruction in the mindfulness intervention used in BASICS-Mobile, called urge surfing,
aims to help the individual to ``ride out'' urges, by recognizing the urge as it arises and allowing the urge to pass on its own. Questions related to these effects concern the proximal effect of treatment on the response defined by
\begin{equation}\label{eq:prox}
  \E\left[Y_{t+1}(\bar A_{t-1}, 1) - Y_{t+1}(\bar A_{t-1}, 0)
  \mid S_{1t}(\bar A_{t-1})\right],
\end{equation}
where $S_{1t}(\bar A_{t-1})$ is a vector of summary variables chosen from $H_t(\bar A_{t-1})$.  The difference in \cref{eq:prox} represents the effect of $A_t = 1$ versus $A_t = 0$ on the response at $t + 1$, given  $S_{1t}(\bar A_{t-1})$.  In conditioning only on $S_{1t}(\bar A_{t-1})$ as opposed to $H_t(\bar A_{t-1})$, the effect \cref{eq:prox} is marginalized over variables in $H_t(\bar A_{t-1})$ that are not in $S_{1t}(\bar A_{t-1})$.  Different choices of variables in $S_{1t}$ address a variety of scientific questions, each of which is useful for understanding the effect of $A_t = 1$ versus $A_t = 0$ on the response $Y_{t+1}$.  For example, a first analysis may focus on the proximal effect that is marginal over all variables in $H_t(\bar A_{t-1})$ (i.e., $S_{1t} = \emptyset$), whereas a second analysis may focus on assessing this effect conditional on particular variables from $H_t(\bar A_{t-1})$.

Note that, for any $A_u$  not contained in $S_{1t}(\bar A_{t-1})$, the expectation in \cref{eq:prox} depends on distribution of $A_u$.  This is a departure from the causal inference literature, where estimands do not  depend on the treatment distribution in the data at hand.  Nonetheless, for all choices of variables in $S_{1t}(\bar A_{t-1})$, the proximal treatment effect is causal, since \cref{eq:prox} is the conditional mean of the contrast between the potential proximal response had an individual received ($a_t = 1$) versus not received ($a_t = 0$) treatment at occasion $t$.  Considering the dependence of the proximal effect on the distribution of the treatments, it is best to always present this distribution along with the estimated treatment effect.  For further discussion concerning including the treatment distribution as part of the estimand, see \cref{sec:discuss}.

Many treatments may have delayed effects.  For example, mindfulness messages have a delayed effect when individuals recall and employ mindfulness exercises provided prior to the most recent treatment occasion.  In BASICS-Mobile, treatments suggesting alternative activities to smoking and drinking may achieve little to no immediate impact in the afternoon, but the individual might follow these suggestions later on in the evening.  So in general both proximal and other lagged effects of treatments on the response variable may be of interest.  To define these lagged effects, we denote $A_{t+1}(\bar A_{t-1}, a)$ by $A_{t+1}^{a_t=a}$, $A_{t+2}(\bar A_{t-1}, a, A_{t+1}^{a_t=a})$ by $A_{t+2}^{a_t=a}$ and so on, with $A_{t+k-1}(\bar A_{t-1},a, A_{t+1}^{a_t},\cdots, A_{t+k-2}^{a_t=a})$ by $A_{t+k-1}^{a_t=a}$.  We define the lag $k$ effect of treatment on the response $k$ treatment occasions into the future $Y_{t+k}$ by
\begin{equation}\label{eq:lagk}
  \E\left[Y_{t+k}(\bar A_{t-1}, 1, A_{t+1}^{a_t=1}, \ldots, A_{t+k-1}^{a_t=1})
  - Y_{t+k}(\bar A_{t-1}, 0, A_{t+1}^{a_t=0}, \ldots, A_{t+k-1}^{a_t=0})
  \Bigmid S_{kt}(\bar A_{t-1})\right],
\end{equation}
where $k$ ranges from $1$ up to the number of lags of scientific interest.  So the proximal effect \cref{eq:prox} corresponds to the lag $k = 1$ treatment effect.  
Note that both future actions, as well as $Y_{t+k}$, depend on treatment at occasion $t$ as emphasized by the superscripts $a_t = 1$ or $a_t = 0$.  
As with \cref{eq:prox}, $S_{kt}(\bar A_{t-1})$ is a vector of variables from the history $H_t(\bar A_{t-1})$.  The lagged effect is also similarly averaged over the conditional distribution of variables in the history $H_t(\bar A_{t-1})$ not represented in $S_{kt}(\bar A_{t-1})$, which might include past treatment or underlying moderators.  In addition, \cref{eq:lagk} is averaged over the distribution of treatments after occasion $t$ but before response $Y_{t+k}$---namely $A_{t+1}^{a_t=a}, \ldots, A_{t+k-1}^{a_t=a}$ for either $a = 1$ or $a = 0$.

The causal effect in \cref{eq:lagk} is a generalization of the treatment ``blip''  in the SNMM.  In SNMMs, the $t$th treatment blip or intermediate effect on $Y_{t+k}$ is usually defined with $S_{kt}(\bar A_{t-1}) = H_{kt}(\bar A_{t-1})$ and with respect to a prespecified future (after time $t$) ``reference'' treatment regime that defines the distribution for $A_{t+1}, \ldots, A_{t+k-1}$.  For example, if we were studying treatment discontinuation, we might have chosen the reference regime $A_u = 0$ for $u > t$, with probability one \citep[cf.][Section~3a]{robins1994}.  In this case the lag $k$ treatment effect \cref{eq:lagk} represents the impact of one last additional treatment on the proximal response $k$ time units later.  The reference treatment regime reflected in \cref{eq:lagk}, however, assigns treatment with probabilities between zero and one and corresponds to the distribution of treatments in the data we have at hand.  For further discussion of the connection between the causal effects defined here and the SNMM, see \cref{app:snmm:defn}.

We now express the proximal and other lagged effects in terms of the observed data.  For this we  assume positivity, consistency and sequential ignorability \citep{robins1994,robins1997a}:
\begin{itemize}
\item Consistency: The observed data $(Y_2,X_2,A_2, \ldots, Y_T, X_T, A_T, Y_{T+1})$ are equal to the potential outcomes as follows:  $Y_2=Y_2(A_1), X_2=X_2(A_1),A_2=A_2(A_1)$ and for each subsequent $t \le T$,  $Y_{t}=Y_{t}(\bar A_{t-1}),X_{t}=X_{t}(\bar A_{t-1}), A_{t}=A_{t}(\bar A_{t-1})$ and lastly $Y_{T+1}=Y_{T+1}(\bar A_{T})$.
\item Positivity: If the joint density at  $\{H_t = h_t, {A}_{t} = {a}_{t}\}$ is greater than zero, then $\pr(A_t = a_t \mid H_t = h_t ) > 0$, almost everywhere.
\item Sequential ignorability: For each $t \leq T$, the potential outcomes $\mathlist{\{Y_{t+1}(\bar a_t), X_{t+1}(\bar a_t), A_{t+1}(\bar a_t),\ldots, Y_{T+1}(\bar a_{T})\}}$ are independent of $A_t$ conditional on $H_t$.
\end{itemize}

The consistency assumption  connects the potential outcomes with the data.   When the treatment allocated to one individual may influence the response of others, the observed response $Y_{t+1}$ is generally consistent not with the potential response $Y_{t+1}(\bar A_t)$ as above, but possibly with some other group-based conceptualization \citep[e.g.][]{hong2006,vanderweele2013}.  In particular, for a mobile intervention with a social media component, it may be necessary to define the potential outcomes for a given individual as a function of the treatments that are provided to individuals in their social network.

In an MRT, treatment is sequentially randomized according to known treatment probabilities, say $\pr(A_t = 1 \mid H_t) =  p_t(1 \mid H_t)$, $t = 1, \ldots, T$, and thus sequential ignorability is ensured by design.  In an observational study, where treatment status is observed rather than randomized, sequential ignorability  is often assumed.  Here the underlying treatment probabilities $  p_t(1 \mid H_t)$, $t = 1, \ldots, T$, are unknown.

In \cref{app:snmm:obs} we show that, under these assumptions, the lag $k$ treatment effect can be expressed in terms of the observed data as
\begin{multline}\label{eq:lagk:obs}
  \E\left[Y_{t+k}(\bar A_{t-1}, 1, A_{t+1}^{a_t=1}, \ldots, A_{t+k-1}^{a_t=1})
  - Y_{t+k}(\bar A_{t-1}, 0, A_{t+1}^{a_t=0}, \ldots, A_{t+k-1}^{a_t=0})
  \bigmid S_{kt}(\bar A_{t-1})\right] \\
  = \E\left[\E[Y_{t+k} \mid A_t = 1, H_t]
    - \E[Y_{t+k} \mid A_t = 0, H_t] \mid S_{kt}\right] \\
  = \E\left[\frac{\indic(A_t = 1) Y_{t+k}}{ p_t(1 \mid H_t)}
    - \frac{\indic(A_t = 0) Y_{t+k}}{1 -  p_t(1 \mid H_t)}
    \Bigmid S_{kt}\right],
\end{multline}
for $t = 1, \ldots, T - k + 1$, respectively.  Note that if $S_{kt} = H_t$, then the lag $k$ effect simplifies to
\begin{equation}\label{eq:lagk:obs:hx}
  \E[Y_{t+k} \mid A_t = 1, H_t] - \E[Y_{t+k} \mid A_t = 0, H_t].
\end{equation}

\section{Estimation}
\label{sec:estimate}

In the following we assume a linear model for the treatment effects.  Fortunately, models for the proximal and other lagged treatment effects can in fact be specified separately, since \cref{eq:lagk} for differing lags $k$ do not constrain one another (\citealp{robins1994,robins1997a}; see \cref{app:model}).  Suppose that the following holds.
\begin{assume}
\item\label{con:modelk} Each lag $k$ treatment effect of interest takes the form
\begin{equation}\label{eq:modelk}
  \E\left[\E[Y_{t+k} \mid A_t = 1, H_t]
    - \E[Y_{t+k} \mid A_t = 0, H_t] \mid S_{kt}\right] = f_{kt}(S_{kt})^\tr\beta_k
\end{equation}
where $f_{kt}(s)$ is a $p$-dimensional vector function of $s$ and time $t$.
\end{assume}
The vector, $f_{kt}(S_{kt})$ may include a vector of basis functions in time, for example, for modeling time-varying effects. When $S_{kt} \ne H_t$, (\ref{eq:modelk}) is a marginal model.
 For example, if $S_{kt}=\emptyset$, then  (\ref{eq:modelk}) is $\E\left[\E[Y_{t+k} \mid A_t = 1, H_t]
    - \E[Y_{t+k} \mid A_t = 0, H_t] \right] = f_{kt}^\tr\beta_k$, which is a model for the lag $k$ treatment effects indexed by $t$ but marginal over  $H_t$.


The rest of this paper is devoted to inference on the unknown $p$-dimensional  $\beta_k$.  Throughout we denote the true value of $\beta_k$ by $\beta_k^*$, $n$ represents the number of individuals in the data and  $\eprob h(Z) = \sum_{i=1}^n h(Z_i) / n$ for some function $h$ of the random vector $Z$.  Assume the data comes from an MRT; in this case sequential ignorability is satisfied.  In particular we assume:
\begin{assume}
\item\label{con:rand} Treatment is sequentially randomized with randomization probability $\pr(A_t = 1 \mid H_t) =  p_t(1 \mid H_t)$, for each $t = 1, \ldots, T$.
\end{assume}
Inference concerning $\beta_k$ using data from observational studies in which the treatment is not sequentially randomized can be handled---if the assumption of sequential ignorability holds---by estimating the treatment probability; see \cref{app:large}.

The following, simple, estimation method includes  centering of the treatment indicators and weighting of the estimating function.     The weights allow us to estimate marginal treatment effects, e.g. conditional on $S_{kt}$ instead of $H_t$.  As discussed above this commonly occurs, for example,  when interest lies in the treatment effect of $A_t$ for $S_{kt} =\emptyset$. The weights are ratios of probabilities, with the denominator weight equal to the randomization probability; the numerator probability is arbitrary as long as this probability depends on $H_t$ only via  $S_{kt}$ (the variables in the treatment effect model, (\ref{eq:modelk})).  Denote the numerator probabilities by, $\tilde p_t(a|S_{kt})$   for $t = 1, \ldots, T$.  The weight at occasion $t$ is $W_t=\frac{\tilde p_t(A_t|S_{kt})}{ p_t(A_t|H_t)}$.

The centering produces orthogonality between estimation of the $\beta_k$ parameter in the treatment effect, $ f_{kt}(S_{kt})^\tr\beta_k$ and estimation of the parameters in a nuisance function.    That is, the method below  will provide a consistent estimator of the  lag $k$ effect even when the nuisance function  $\E[W_tY_{t+k} \mid H_t]$ is misspecified.   This robustness property is desirable for two reasons.  First,  the history $H_t$ is usually high dimensional, making it very difficult to model these nuisance functions correctly.  Second, even when $H_t$ is not very large, it can be difficult or impossible to specify models that can be correct for both the nuisance function as well as for the delayed treatment effects at lags $j > k$ (see \cref{app:model} for an example).    
Below we provide results when the working model for $\E[W_tY_{t+k} \mid H_t]$ is $g_{kt}(H_t)^\tr\alpha_k$ where $g_{kt}(H_t)$ is a vector of features constructed from $H_t$ and the vector $\alpha_k$ is unknown.

  The centered and weighted least squares estimating function is
\begin{eqnarray}\label{eq:lsw}
  U_{\mathrm{W}}(\alpha_k, \beta_k) &=& \sum_{t=1}^{T-k+1}
  \big(Y_{t+k} - g_{kt}(H_t)^\tr\alpha_k -  (A_t - \tilde p_t(1 \mid S_{kt})) f_{kt}(S_{kt})^\tr\beta_k\big)\phantom{bbbbbbbbbbbb}  \nonumber\\
& &\phantom{bbbbbbbbbbbbbbbbbbbbbbbbbbbbbbbbbbbbb}  W_t\begin{pmatrix} g_{kt}(H_t) \\ (A_t - \tilde p_t(1 \mid S_{kt})) f_{kt}(S_{kt}) \end{pmatrix},
\end{eqnarray}
where as before, $W_t=\frac{\tilde p_t(A_t|S_{kt})}{ p_t(A_t|H_t)}$.
  Let $\dot{U}_{\mathrm{W}}$ be the derivative of $U_{\mathrm{W}}$ with respect to the row vector $(\alpha_k^\tr, \beta_k^\tr)$. In \cref{app:large} we prove a more general version of the following result.

\begin{proposition}\label{thm:weight}
Assume  \cref{con:modelk} and \cref{con:rand}, both defined above.  Then, under invertibility and moment conditions, the solution to the estimating equation $\eprob U_{\mathrm{W}}(\alpha_k, \beta_k) = 0$ yields an estimator $(\hat{\alpha}_k, \hat{\beta}_k)$ for which $\sqrt{n}(\hat{\beta}_k - \beta_k^*)$ is asymptotically normal with mean zero and variance-covariance matrix consistently estimated by the lower block diagonal ($p \times p$) entry of the matrix $(\eprob \dot{U}_{\mathrm{W}}(\hat{\alpha}_k, \hat{\beta}_k))^{-1} \eprob U_{\mathrm{W}}(\hat{\alpha}_k, \hat{\beta}_k)^{\otimes 2} {(\eprob \dot{U}_{\mathrm{W}}(\hat{\alpha}_k, \hat{\beta}_k))^{-1}}^\tr$.
\end{proposition}

\noindent\underbar{Remarks} 
\begin{enumerate}
\item  A first look at the estimating function, (\ref{eq:lsw}), might lead one to think that the estimating function is unbiased only if  $\E[Y_{t+k} \mid A_t, H_t]=g_{kt}(H_t)^\tr\alpha_k +  (A_t - \tilde p_t(1 \mid S_{kt})) S_{kt}^\tr\beta_k $ for some $(\alpha_k,\beta_k)$; however this is not the case.   Indeed,  the primary assumption \cref{con:modelk} only concerns a marginal quantity derived from $\E[Y_{t+k} \mid A_t, H_t]$.  Furthermore,  the working model $g_{kt}(H_t)^\tr\alpha_k$ for $\E[W_tY_{t+k} \mid  H_t]$ need not be correct in order for $\hat{\beta}_k$ to be consistent and for the large sample results to hold (see the proof in \cref{app:large}).
\item  The numerator of the weight can be set to the denominator (the randomization probability) and thus the weight will be $1$ if the randomization probabilities depend at most on $S_{kt}$, $p_t(\cdot|H_t)=p_t(\cdot|S_{kt})$; here, choose $\tilde p_t(\cdot|S_{kt})=p_t(\cdot|S_{kt})$ so that $W_t=1$.  Furthermore, if the randomization probabilities are constant, $\rho$, then setting  $\tilde p_t( 1 |S_{kt})=\rho$, simplifies (\ref{eq:lsw}) to an unweighted regression with recoded treatment indicators ($A_t\to A_t-\rho$).

\item The weight $W_t$ is reminiscent of inverse probability of treatment weighting in causal inference \citep{robins1998}.  However, in addition to facilitating estimation of marginal treatment effects, here weighting (and centering) is simply used to make the weighted least squares 
estimator $\hat{\beta}_k$ robust against the  case in which the working model $g_{kt}(H_t)^\tr\alpha_k$ misspecifies $\E[W_tY_{t+k} \mid  H_t]$. Further, this  similarity might lead one to use the numerator of the weight to ``stabilize'' the weights \citep[e.g. Section 6.1 of][]{robins2000b}; that is, to select a $\tilde p_t$ to make $W_t$ as close to $1$ as possible. There are two caveats to this.  First, the numerator probabilities determine the limit of $\hat\beta_k$ when the modeling assumption for the lag $k$ treatment effect (\ref{eq:modelk}) is false and thus might be selected with this alternative in mind;  see (\ref{eq:projbeta}, \ref{eq:projbeta1}) in  \cref{app:large}.  Second, bias can result if the numerator of the weight depends on variables that are not in $S_{kt}$;  see the second simulation in 
\cref{sec:sim}.
 
\item Centering has been previously employed by \citet{brumback2003} and \citet{goetgeluk2008} for causal inference. For example \citet{goetgeluk2008} center exposure variables by their overall mean to protect against unmeasured baseline confounders.  \citet{brumback2003} center time-varying exposures by their conditional mean given the history, as we do; they consider treatment effects under a treatment discontinuation reference regime and limit attention to overall effects without interaction terms.   In contrast to these papers, our use of centering is similar to that of \citepos{liao2015} and  is solely to  provide robustness to the working  model for $\E[W_tY_{t+k} \mid H_t]$; centering is not used to adjust for confounding.  In  \citealp{liao2015} the treatment probabilities are non-stochastic.
\item The similarity of (\ref{eq:lsw}) to generalized estimating equations (GEEs,  \citealp{liang1986}) might motivate the inclusion of a non-independence working correlation matrices such as exchangeable or AR(1) in the estimating function so as to reduce variance of $\hat\beta_k$ \citep[e.g.][]{mancl1996}.  
   Similarly, an analyst might wish to use a non-independence working correlation matrix in our setting for the same reason, but this strategy will generally introduce bias.  Such a result is unsurprising given the bias that arises when non-independence working matrices are used  in  inverse probability of treatment weighting literature \citep{vansteelandt2007,tchetgen2012} or in GEEs where a time-varying response is modeled by time-varying covariates \citep{pepe1994}.  
   The simulations in Table~\ref{tab:sim:ar1} in Section~\ref{sec:sim}, and Table~\ref{tab:sim:ar1:supplement} in \cref{app:sim}  illustrate such bias. 
\end{enumerate}

\section{Availability}
\label{sec:avail}

Up to this point we have implicitly presumed that at every possible occasion $t$, the participant is available to engage with the mobile intervention.  Consideration of availability is critical  since it might be unreasonable, counter-productive or even unethical to always presume availability.  By experimental design, treatment will not be delivered to unavailable individuals.  For example in HeartSteps \citep{klasjna2015}, smartphone notifications are used to deliver suggestions to disrupt sedentary behavior.  Here the participant is considered unavailable when driving a vehicle (because the notification may be distracting) or walking (as treatment at this time is scientifically inappropriate).  Detection of availability can be carried out through sensors (as in the case of HeartSteps) or recent interaction with the mobile device.  BASICS-Mobile took the latter approach by presuming that participants were available to receive a treatment only after they fully completed a self-report.

Assume that the measurements $X_t$ just prior to the $t$th treatment occasion contain the participant's availability status, denoted by $I_t$, where $I_t = 1$ if the participant is available to engage with the treatment at occasion $t$ and $I_t = 0$ otherwise.    To define the treatment effects under limited availability, we use potential outcome notation. The potential outcome notation allows us to not only make explicit the dependence of $Y_{t+1}$ on treatment $\bar a_t$ but also make explicit the dependence of $I_t$ on $\bar a_{t-1}$.   Furthermore, in contrast to \cref{sec:effects}, here the potential outcomes are indexed by decision rules because treatment can only be provided when a participant is available.   The use of decision rules to index potential outcomes helps make explicit that, by  experimental  design,  treatment $A_t$ is not delivered if the participant is unavailable at the  $t$ treatment occasion.  
 In particular define $d(a, i)$ for $a \in \{0, 1\}$, $i \in \{0, 1\}$ by $d(a, 0) = 0$ and $d(a, 1) = a$  (recall that here $a=0$ means no treatment).  Then for each ${a}_1 \in \{0, 1\}$, define $D_1(a_1) = d(a_1, I_1)$.  The potential proximal responses  following treatment occasion $1$ are  $\{Y_2(D_1(1)), Y_2(D_1(0))\}$.   Note that if $I_1=0$ then $D_1(1)=D_1(0)=0$ and thus $\{Y_2(D_1(1)), Y_2(D_1(0))\}= \{Y_2(0), Y_2(0)\}$.  That is, the experimental design excludes the possibility to observe $Y_2(1)$ if $I_1=0$.   Similarly there are  potential outcomes for availability; this emphasizes the fact that previous exposure to treatment can influence subsequent availability.   In BASICS-Mobile, for example, repeated provision of treatment might lead to lower engagement with the intervention, and therefore lower availability for further delivery of the treatment.  The  potential availability indicators at  $t=2$ are $\{I_2(D_1(1)), I_2(D_1(0))\}$.   As with the proximal response, if $I_1=0$ then $D_1(1)=D_1(0)=0$ and thus $\{I_2(D_1(1)), I_2(D_1(0))\}= \{I_2(0), I_2(0)\}$.

The decision rules at $t>1$ are defined iteratively, building on prior decision rules.  For each $\bar a_2 = (a_1, a_2)$ with $a_1, a_2 \in \{0, 1\}$, define $D_2(\bar a_2) = d(a_2, I_2(D_{1}(a_{1})))$ and $\overline{D_2(\bar a_2)} = (D_1(a_1), D_2(\bar a_2))$.   A potential  proximal response following occasion $t=2$ and corresponding to $\bar a_2$ is  $Y_3(\overline{D_{2}(\bar a_{2})})$ and a potential  availability indicator at $t=3$ is  $I_3(\overline{D_{2}(\bar a_{2})})$.  Similarly, for each $\bar a_t=(a_1, \dots, a_t) \in   \{0, 1\}^t$, define $D_t(\bar a_t) = d(a_t, I_t(\overline{D_{t-1}(\bar a_{t-1})}))$ and $\overline{D_t(\bar a_t)} = (D_1(a_1), \ldots, D_t(\bar a_t))$.   For each $\bar a_t = (a_1, \dots, a_t) \in \{0, 1\}^t$, the potential proximal response is  $Y_{t+1}(\overline{D_{t}(\bar a_{t})})$ and potential  availability indicator is  $I_{t+1}(\overline{D_{t}(\bar a_{t})})$ at occasion $t+1$.

We now incorporate availability into the definition of the proximal treatment effect;  first recall the notation from the end of \cref{sec:notation}; similarly denote $A_2(D_1(A_1))$ by $A_2$ and so on with $A_t(\overline{D_{t-1}(\bar A_{t-1})})$ denoted by $A_t$.
 The proximal treatment effect is
\begin{equation*}
  \E\left[Y_{t+1}\left(\overline{D_t(\bar A_{t-1}, 1)}\right)
    - Y_{t+1}\left(\overline{D_t(\bar A_{t-1}, 0)}\right)
    \bigmid I_t\left(\overline{D_{t-1}(\bar A_{t-1})}\right) = 1,
    S_{1t}\left(\overline{D_{t-1}(\bar A_{t-1})}\right)\right].
\end{equation*}
Unlike \cref{eq:prox}, this effect is defined for only individuals available for treatment at time $t$, that is, $I_t\left(\overline{D_{t-1}(\bar A_{t-1})}\right) = 1$.  This subpopulation is not static; at a given treatment occasion $t$ only certain types of individuals might tend to be available and availability for any given individual may change with $t$.  Conditioning on availability is related to the concept of viable or feasible dynamic treatment regimes \citep{wang2012,robins2004}, in which one assesses only the causal effect of treatments that can actually be provided.

To incorporate availability into the definition of the lagged effects, we use the shorthand notation: denote $A_{t+1}(\overline{D_t(\bar A_{t-1},a)})$ by $A_{t+1}^{a_t=a}$, $A_{t+2}(\overline{D_{t+1}(\bar A_{t-1},a)}, A_{t+1}^{a_t=a})$ by $A_{t+2}^{a_t=a}$, and so on, with $A_{t+k-1}(\overline{D_{t+1}(\bar A_{t-1},a)}, A_{t+1}^{a_t},\cdots, A_{t+k-2}^{a_t=a})$ by $A_{t+k-1}^{a_t=a}$.
The lag $k$ effect of treatment on the response $k$ treatment occasions into the future $Y_{t+k}$ is defined by
\begin{multline*}
  \E\Bigl[Y_{t+k}\left(\overline{D_t(\bar A_{t-1}, 1)}, A_{t+1}^{a_t=1},
        \ldots, A_{t+k-1}^{a_t=1}\right) \\
    - Y_{t+k}\left(\overline{D_t(\bar A_{t-1}, 0)}, A_{t+1}^{a_t=0},
        \ldots, A_{t+k-1}^{a_t=0}\right)
    \bigmid S_{kt}\left(\overline{D_{t-1}(\bar A_{t-1})}\right)\Bigr].
\end{multline*}

Assuming consistency, positivity and sequential ignorability, the lag $k$ treatment effect under limited availability can be expressed in terms of the data as
\begin{multline*}
  \E\left[\E[Y_{t+k} \mid A_t = 1, I_t = 1, H_t]
    - \E[Y_{t+k} \mid A_t = 0, I_t = 1, H_t] \mid I_t = 1, S_{kt}\right] \\
  = \E\left[\frac{\indic(A_t = 1) Y_{t+1}}{ p_t(1 \mid H_t)}
    - \frac{\indic(A_t = 0) Y_{t+1}}{1 -  p_t(1 \mid H_t)}
    \Bigmid I_t = 1, S_{kt} \right],
\end{multline*}
where $ p_t(1 \mid H_t)$ is now $\pr(A_t = 1 \mid I_t = 1, H_t)$.  Modeling and estimation proceeds following the same approach as with the always-available setting.  In particular for the lag $k$ treatment effect, we assume the linear model
\begin{equation}
\label{eq:withavail}
  \E\left[\E[Y_{t+k} \mid A_t = 1, I_t = 1, H_t]
    - \E[Y_{t+k} \mid A_t = 0, I_t = 1, H_t] \mid I_t = 1, S_{kt}\right]
  = f_{kt}(S_{kt})^\tr\beta_k,
\end{equation}
where, as before, $f_{kt}(S_{kt})$ is a vector of features involving $S_{kt}$ and time $t$. To form the estimating function for $\beta_k$, we replace $W_t$ in (\ref{eq:lsw}) by the product $I_tW_t$. The working model and the treatment probability models are conditional on $I_t = 1$. A more general version of the resulting estimating equation is provided in display \cref{eq:geew} of \cref{app:large}.  Proofs can be found in \cref{app:large}.

\section{Implementation}
\label{sec:implement}

The weighting and centering estimation method can be implemented using standard software for GEEs, provided that we: (i) incorporate $I_t W_t$ as ``prior weights'' and (ii) employ a independence working correlation matrix.
The standard errors provided in \cref{thm:weight} directly correspond to the sandwich variance-covariance estimator provided by GEE software.  From existing work on GEEs, it is well understood that the sandwich estimator is non-conservative in small samples.  To address this, whenever $n \leq 50$, we apply \citepos{mancl2001} small sample correction to the term $\eprob U_{\mathrm{W}}(\hat{\alpha}_k, \hat{\beta}_k)^{\otimes 2}$ in the estimator of the variance; in particular we premultiply the $(T-k+1) \times 1$ vector of each person's residuals in  $U_{\mathrm{W}}$ by the inverse of the identity matrix  minus the leverage for this person.   Also, as in  \citet{liao2015}, we use critical values from a $t$ distribution or a Hotelling's T-squared distribution.  In particular if we wish to test the null hypothesis for a linear combination of $\beta_k$---e.g., test $c^{\tr}\beta_k=0$ for a known $p$-dimensional vector $c$---then we use the critical value  $t_{n-p-q}^{-1}(1-\alpha_0)$ where, $p$ is the dimension of $\beta_k$, $q$ is the dimension of $\alpha_k$ and $\alpha_0$ is the significance level.  More generally, if we wish to conduct a $p'$-dimensional multivariate test of $\beta_k$---e.g., test $z^{\tr}\beta_k=0$ for a known $p \times p'$ matrix $z$---then the critical value is $F_{p', n-q-p}^{-1}\left(\frac{(n-q-p')(1-\alpha_0)}{p'(n-q-1)}\right)$. 
 
 %

When either  $\tilde p_t(1 \mid S_{kt})$ or $ p_t(1 \mid H_t)$ is estimated, the sandwich variance-covariance estimator must be adjusted to account for the additional sampling error (see \cref{app:large}).  See \cref{app:code} to obtain code that calculates standard errors using R \citep{r2015}. 

\section{Simulation Study}
\label{sec:sim}

Here, we evaluate the proposed centering and weighting method via simulation experiments.   

%
%
%
%
%

The following, simple, generative model will allow us to illustrate the  proposed method and compare it with existing methods.  Consider data arising from an MRT (so the randomization probability $ p_t(1|H_t)$ is known).  The generative model for the response, $Y_{t+1}$, is a linear model in $(A_t, S_t, A_{t-1}, S_{t-1}, A_{t-2}, A_tS_t, A_{t-1}S_t, A_{t-2}S_{t-1})$, for $S_t \in \{-1, 1\}$.  For  convenience in reading off the marginal effects, we write this model as $Y_{t+1} = \theta_1 (S_t - \E[S_t \mid A_{t-1}, H_{t-1}]) + \theta_2(A_{t-1} - p_{t-1}(1 \mid H_{t-1})) +  (A_t -  p_t(1 \mid H_t)) (\beta_{10}^* + \beta_{11}^* S_t) + \epsilon_{t+1}$.  Here the randomization probability is given by $ p_t(1 \mid H_t) = \expit(\eta_1 A_{t-1} + \eta_2 S_t)$, $\pr(S_t = 1 \mid A_{t-1}, H_{t-1}) = \expit(\xi A_{t-1})$ (note $A_0=0$), and $\epsilon_t \sim \norm(0, 1)$ with $\cor(\epsilon_u, \epsilon_t) = 0.5^{|u - t|/2}$.   Throughout each subject is available at every treatment occasion: $I_t = 1$ ($t = 1, \ldots, T$).  In the simulation scenarios below, we fix $\theta_1=0.8$ and $\beta_{10}^* = -0.2$ and we vary $(\theta_2$, $\beta_{11}^*, \eta_1, \eta_2, \xi)$.

The marginal proximal (lag $k=1$) effect is given by $\E[\E[Y_{t+1} \mid A_t = 1, H_t] - \E[Y_{t+1} \mid A_t = 0, H_t] ] = \beta_{10}^* + \beta_{11}^* \E[S_t]$.   Note that if $\beta^*_{11}=0$ or $E[S_t]=0$ (i.e., by setting $\xi = 0$), then the marginal proximal treatment effect is constant in time and is given by $\beta_1^*=\beta_{10}^*=-0.2$. 

Here, we consider three simulation experiments.  All three simulation experiments concern estimation of the marginal proximal treatment effect $\beta_1$. Thus in all cases when the weighted and centered method is used, $f_{1t}(S_{1t})=(1)$ in the estimating function (\ref{eq:lsw}) (i.e., $S_{1t}=\emptyset$).  
We report average average $\hat\beta_1$ point estimates, standard deviation and root mean squared error of $\hat\beta_1$, and 95\% confidence interval coverage probabilities for $n = T = 30$ across 1000 replicates.  Confidence intervals are  based on  standard errors that are corrected for the estimation of weights and/or small samples (see \cref{sec:implement}).  
The tables below omit the average estimated standard errors; these are provided in \cref{app:sim} and closely correspond to the standard deviations of the point estimates.
\cref{app:sim} also reports additional results for $n=30, 60$ with $T=30, 50$ (results were similar for different $T$ values), and compares the proposed method versus centering but not weighting ($W_t=1$ for all $t$) in a fourth simulation experiment.


The first simulation experiment concerns the estimation of $\beta^*_1$ when an important moderator exists. This experiment illustrates that, when primary interest is in the marginal proximal treatment effect, weighting and centering is preferable over GEE. 
In the data generative model, we set $\theta_2 = 0$, $\eta_1 = -0.8$, $\eta_2 = 0.8$ 
and $\xi = 0$  (recall $\xi=0$ implies that the true marginal proximal treatment effect is $\beta^*_1=-0.2$).  Different scenarios were devised by setting $\beta_{11}^*$ to one of $0.2$, $0.5$, $0.8$, giving respectively a small, medium, or large degree of moderation by $S_t$.  Since $\eta_1$ and $\eta_2$ are nonzero, the treatment $A_t$ is assigned with a probability depending on both $S_t$ and past treatment $A_{t-1}$, for each $t$.

In the weighted and centered analysis, we parameterize and estimate $\tilde{p}_t$.   In particular,  $\tilde{p}_t(a;\hat\rho)=\hat\rho^a(1-\hat\rho)^{1-a}$ where  $\hat\rho = \eprob\sum_{t=1}^T A_t/T$.  The weights are set to $W_t = {\hat\rho}^{A_t}(1 - \hat\rho)^{1-A_t}/ p_t(A_t \mid H_t)$ and the working model for $\E[W_tY_{t+1} \mid H_t]$ is $\alpha_{10} + \alpha_{11} S_t$ (i.e., $g_{1t}(H_t)=(1,S_t)^\tr$).  Thus the estimating function in (\ref{eq:lsw}) is given by
\begin{eqnarray*}
 \sum_{t=1}^{T}
\big(Y_{t+1} - (\alpha_{10} + \alpha_{11} S_t) -  (A_t - \hat\rho) \beta_1\big) W_t\begin{pmatrix} (1,S_t)^\tr \\ A_t - \hat\rho \end{pmatrix}.
\end{eqnarray*}


A common alternative would be a GEE analysis with an independence working correlation matrix.   The GEE estimating function with an independence working correlation matrix (GEE-IND)  is the above  estimating function but with $W_t=1$ for all $t$ and $A_t$ not centered.  A more likely alternate  that would be used in the mobile health literature is a GEE with an non-independence working correlation matrix \citep{schafer2006}; the resulting conditional mean model is the same as when random effects are used \citep{Stone2007, bolger2013}. We also provide a comparison with this alternative, using an AR(1) correlation matrix (GEE-AR(1)).   Note that, to guarantee consistency in a GEE analysis, one would assume that the  analysis model is correct; since here the analysis model is 
$Y_{t+1}\sim\alpha_{10} + \alpha_{11} S_t + A_t \beta_1$, the corresponding assumption would be that  $\E[Y_{t+1}\mid S_t, A_t]= \alpha_{10} + \alpha_{11} S_t + A_t \beta_1$ for some $(\alpha_{10},\alpha_{11},\beta_1)$.  This assumption is false (no $A_tS_t$ term).  The weighting and centering method, on the other hand, does not require a model for the conditional mean.   For consistency, the weighting and centering method only uses the assumption that  $\E\big[\E[Y_{t+1}\mid S_t, A_t=1]-\E[Y_{t+1}\mid S_t, A_t=0]\big]=\beta_1$ for some $\beta_1$.

Since the treatment effect term does not include $S_t$, the GEE conditional mean models are misspecified.  Furthermore since $\eta_2 = 0.8$, the randomization probability $ p_t(1 \mid H_t)$ depends on the underlying moderator $S_t$.  We therefore anticipate the $\hat\beta_{1}$ from the GEE methods to be a biased estimator of the marginal treatment effect of $\beta^*_1=-0.2$ and we expect this bias to increase proportional to $\beta_{11}^*$.   On the other hand, all of the requirements needed to achieve consistency in the proposed method  are satisfied; hence, the $\hat\beta_{1}$ from the weighted and centered method should be unbiased, regardless of the value for $\beta_{11}^*$.   These conjectures concerning bias are supported by \cref{tab:sim:omit}.  In addition, (i) for $\beta^*_{11}=0.5, 0.8$ the RMSE for GEE is greater than or equal to the RMSE for the proposed method; and (ii) for all $\beta^*_{11}$ the proposed method achieves nominal $95$\% coverage, whereas, the GEE methods generally do not (an exception was for $\beta^*_{11}=0.2$ with GEE-IND).  For further results see \cref{tab:sim:omit:supplement} in the Supplement.

\begin{table}[H]
\caption{Comparison of three estimators of the marginal proximal treatment effect, $\hat\beta_1$, when an important moderator is omitted.}
\label{tab:sim:omit}
{\centering
\begin{tabular}{l cccc cccc cccc}
	\toprule
	& \multicolumn{4}{c}{Weighted and Centered} & \multicolumn{4}{c}{GEE-IND} & \multicolumn{4}{c}{GEE-AR(1)} \\
	\cmidrule(lr){ 2-5 }\cmidrule(lr){ 6-9 }\cmidrule(lr){ 10-13 } 
	$\beta^*_{11}$ & Mean & SD & RMSE & CP &        Mean & SD & RMSE & CP                & Mean & SD & RMSE & CP \\ 
	\midrule
	$0.2$          & --0.20 & 0.08 & 0.08 & 0.96  & \bftab --0.17 & 0.07 & 0.07 &        0.94 & \bftab --0.16 & 0.04 & 0.06 & \bftab 0.86 \\ 
	$0.5$          & --0.20 & 0.08 & 0.08 & 0.95  & \bftab --0.14 & 0.07 & 0.09 & \bftab 0.88 & \bftab --0.13 & 0.05 & 0.09 & \bftab 0.70 \\ 
	$0.8$          & --0.20 & 0.08 & 0.08 & 0.95  & \bftab --0.10 & 0.07 & 0.12 & \bftab 0.78 & \bftab --0.10 & 0.05 & 0.12 & \bftab 0.57 \\ 
	\bottomrule
\end{tabular}
\par}\smallskip
\begin{minipage}{0.95\linewidth}
RMSE, root mean squared error and SD, standard deviation of $\hat\beta_1$; CP, $95$\% confidence interval coverage probability for $\beta_1^*=-0.2$.  Results are based on $1000$ replicates with $n = T = 30$. Boldface indicates whether Mean or CP are significantly different, at the $5$\% level, from $-0.2$ or $0.95$, respectively.  GEE-IND is the same as the proposed method but with $W_t=1$ and no centering. In GEE-AR(1) includes an AR(1) working correlation matrix.
\end{minipage}
\end{table}



The second and third simulation experiments focus on the proposed weighted and centered estimator. 
The second experiment illustrates that the ability to stabilize the weights is limited, since weighted least squares is prone to bias if the numerator of $W_t$ depends on variables that are not in $S_{kt}$. 
In the data generative model, we set $\theta_2 = -0.1$, $\beta_{11}^* = 0.5$, $\eta_1 = -0.8$, $\eta_2 = 0.8$ and $\xi = 0$.  Thus as above, the randomization probability for $A_t$ depends on both $S_t$ and past treatment $A_{t-1}$ ($t = 1, \ldots, T = 100$).  Here, since $\beta_{11}^* = 0.5$, $S_t$ is a moderator of the proximal effect of treatment
and  since $\theta_2=\beta^*_1/2=-0.1$ there is a lag $k=2$ treatment effect of $A_{t-1}$ on $Y_{t+1}$.

In the data analysis using (\ref{eq:lsw}), the weighted and centered method,  the working model for $\E[W_tY_{t+1} \mid H_t]$  is again $\alpha_{10} + \alpha_{11}S_t$; thus, $g_{1t}(H_t)=(1,S_t)$. 
As before we assume $\E\big[\E[Y_{t+1}\mid S_t, A_t=1]-\E[Y_{t+1}\mid S_t, A_t=0]\big]=\beta_1$ for some $\beta_1$ thus $f_{1t}(S_{1t})=1$.   The denominator of the weight $W_t$ is the known randomization probability, $p_t(A_t \mid H_t)$.  We consider two different choices for $\tilde{p}_t$ (hence, two different choices for centering $A_t$ and for the numerator of $W_t$): 
(i) A choice that is constant in $t$. Here, $\tilde{p}_t(a;\hat\rho) = \hat\rho^a(1-\hat\rho)^{1-a}$ where  $\tilde{p}_t(1;\hat\rho)=\hat\rho = \eprob\sum_{t=1}^T A_t/T$. 
The weights are $W_t(A_t, H_t) = {\hat\rho}^{A_t}(1 - \hat\rho)^{1-A_t}/ p_t(A_t \mid H_t)$; 
(ii) A choice that depends on $S_t$. Here, instead, $\tilde{p}_t(1 \mid S_t;\hat\rho) = \expit(\hat\rho_0 + \hat\rho_1 S_t)$, where $\hat\rho=(\hat\rho_0, \hat\rho_1)$ is the solution to $\eprob \sum_t \exp(\rho_0 + \rho_1 S_t) \{\expit(\rho_0 + \rho_1 S_t) (1 - \expit(\rho_0 + \rho_1 S_t))\}^{-1} (A_t - \expit(\rho_0 + \rho_1 S_t)) (1, S_t)^\tr = 0$.  
In (i) the probability in the numerator is constant for all $W_t$ ($t = 1, \ldots, T=30$).  In (ii) the probability in the  numerator depends on $S_t$ yet interest is in a marginal proximal effect $\beta_1$ ($S_t$ is not a part of $f_{1t}(S_{1t})$).  Hence, we anticipate bias in $\hat\beta_1$ under (ii), but not (i).  This is indeed reflected in \cref{tab:sim:stable},  with (ii) exhibiting bias and achieving a coverage probability of $89$\%.
For further results see \cref{tab:sim:stable:supplement} in \cref{app:sim}. 

\begin{table}[H]
\caption{Weighted and centered estimator of the marginal proximal treatment effect, $\hat\beta_1$, using two choices for $\tilde{p}_t$.}
\label{tab:sim:stable}
{\centering
\begin{tabular}{l cccc}
	\toprule
	$\tilde{p}_t$ & Mean & SD & RMSE & CP \\ 
	\midrule
	Constant in $t$ (i)          &        --0.20  & 0.08 & 0.08 &        0.94 \\ 
	Depends on $S_t$ (ii)        & \bftab --0.14  & 0.09 & 0.11 & \bftab 0.89 \\ 
	\bottomrule
\end{tabular}\par}
\smallskip
\begin{minipage}{0.95\linewidth}
	RMSE, root mean squared error and SD, standard deviation of $\hat\beta_1$; CP, $95$\% confidence interval coverage probability for $\beta_1^*=-0.2$.  Results are based on $1000$ replicates with $n = T = 30$. Boldface indicates whether Mean or CP are significantly different, at the $5$\% level, from $-0.2$ or $0.95$, respectively.
\end{minipage}
\end{table}


The third simulation experiment illustrates that employing a non-independence working correlation structure with the weighted and centered method can result in bias.
In the data generative model, we set $\theta_2 = -0.1$, $\beta_{11}^* = 0$, $\eta_1 = \eta_2 = 0$ and $\xi = 0.1$.  There is no moderation of the proximal effect, since $\beta_{11}^* = 0$.  Unlike the above scenarios, here the  predictor $S_t$ is influenced by $A_{t-1}$ (since $\xi = 0.1$), and because $\theta_2=\beta^*_1/2=-0.1$, there is a lag $k=2$ treatment effect of $A_{t-1}$ on $Y_{t+1}$.  Treatment is randomized with fixed probability $ p_t(1 \mid H_t) = 0.5$ for each $t = 1, \ldots, T = 30$ since $\eta_1=\eta_2=0$.

In the data analysis using (\ref{eq:lsw}), the weighted and centered method,  the working model for $\E[W_tY_{t+1} \mid H_t]$  is again $\alpha_{10} + \alpha_{11}S_t$; thus, $g_{1t}(H_t)=(1,S_t)$. 
In both data analyses, we correctly model $\E\big[\E[Y_{t+1}\mid S_t, A_t=1]-\E[Y_{t+1}\mid S_t, A_t=0]\big]$ by a constant, here denoted by $\beta_1$ thus $f_{1t}(S_{1t})=1$.  We set $\tilde{p}_t(1)=0.5$  thus the weights are $W_t=1$ for all $t=1,\ldots,T=30$. We compare the use of (i) the estimating function in (\ref{eq:lsw}), which corresponds to an independent working correlation structure, versus (ii) using a working AR(1) correlation matrix assuming a correlation of $0.5^{|u - t|/2}$ between times $u$ and $t$.   In the latter case, the estimating function is
\begin{eqnarray*}
 \sum_{t=1}^{T} \begin{pmatrix} (1,S_t)^\tr \\ A_t - 0.5 \end{pmatrix} \sum_{u=1}^T v_{tu} 
\big(Y_{u+1} - (\alpha_{10} + \alpha_{11}S_u)  -  (A_u - 0.5) \beta_1\big),
\end{eqnarray*} 
where $v_{tu}$ is the $(t,u)$ entry of $V^{-1}$, where the $(t,u)$ entry in $V$ is $0.5^{|u - t|/2}$. While AR(1) might better represent the true correlation matrix than an independence correlation matrix, we expect (ii) to induce bias as this marginal model includes time-varying covariates.
 \Cref{tab:sim:ar1} demonstrates this result, with (ii) exhibiting bias and achieving a coverage probability of $65$\%.   Further results are provided in  \cref{tab:sim:ar1:supplement} in the Supplement. 

\begin{table}[H]
\caption{Weighted and centered estimator of the proximal effect, $\hat\beta_1$, with different working correlation structures. }
\label{tab:sim:ar1}
{\centering
\begin{tabular}{lrrrr}
	\toprule
	Working Correlation &      Mean & SD   & RMSE &        CP \\ 
	\midrule
	Independent (i) &        --0.20 & 0.07 & 0.07 &        0.96 \\ 
	AR(1)      (ii) & \bftab --0.13 & 0.06 & 0.09 & \bftab 0.66 \\ 
	\bottomrule
\end{tabular}
\par}
\smallskip
\begin{minipage}{0.95\linewidth}
	RMSE, root mean squared error and SD, standard deviation of $\hat\beta_1$; CP, $95$\% confidence interval coverage probability for $\beta_1^*=-0.2$.  Results are based on $1000$ replicates with $n = T = 30$. Boldface indicates whether Mean or CP are significantly different, at the $5$\% level, from $-0.2$ or $0.95$, respectively.
\end{minipage}
\end{table}

\section{Application}
\label{sec:apply}

BASICS-Mobile is a pilot study, with  $n = 28$, $T = 28$.  The response $Y_{t+1}$ is the smoking rate from the $t$th occasion to the next self-report, and participants are  presumed available only if they completed the preceding self-report.  So the availability $I_t$ is the self-report completion status just prior to $t$ and the treatment decision $D_t$ is $1$ only if a mindfulness message is provided at $t$.  Otherwise, $D_t = 0$.

BASICS-Mobile was neither a sequentially randomized trial nor an observational study. Treatment delivery at occasion $t$ was based on a complex decision rule involving primarily a self-reported measure that the user had an urge or inclination to smoke at the preceding self-report ($\textit{urge}_t$), an indicator for the first three treatment occasions ($\indic(t < 4)$), and a combination of other variables.  For illustrative purposes we provide an analysis acting as though the study was observational and assuming sequential ignorability; we estimate (with logistic regression) the treatment probabilities in the denominator of the weights, $p_t(1 \mid H_t)$, based on $(Y_t, \textit{urge}_t, \indic(t < 4))$ using
\begin{equation*}
   p_t(1 \mid H_t; \hat{\eta})
  = \expit\left(0.69 + 0.02 Y_t + 0.17 \textit{urge}_t - 0.28 \indic(t < 4)
  + 0.70 \textit{urge}_t\indic(t < 4)\right).
\end{equation*}

We examine proximal ($k=1$) and lag-2 ($k=2$) treatment effects. For the proximal effect analysis, we examine one candidate time-varying moderator $S_{1t} = \textit{incr}_t$, which indicates whether or not the user reported an increase in need to self-regulate thoughts over the two self-reports preceding $t$.  
Thus in the estimating function (\ref{eq:lsw}) for the proximal effect analysis, we set $f_{1t}(S_{1t}) = (1, \textit{incr}_t)^\tr$.  For the delayed effect analysis, we consider only the marginal lag-2 effect; thus, $f_{2t}(S_{2t}) = (1)$ in the estimating function (\ref{eq:lsw}). 
For both analyses, we centered and estimated the numerator of the weights based on $\tilde{p}_t(a;\hat\rho)= \hat\rho^a(1-\hat\rho)^{1-a}$ where  $\hat\rho = \eprob\sum_{t=1}^T I_tA_t / \eprob\sum_{t=1}^T I_t = 0.67$. Hence, for both analyses, the weights were set to $W_t= \hat\rho^{A_t}(1-\hat\rho)^{1-A_t} / p_t(A_t \mid H_t; \hat{\eta}) $.
In the working model for both analyses, a variety of predictors are incorporated in $g_{kt}(H_t)$ ($k = 1, 2$), including an intercept term, $\textit{incr}_t$, current urge to smoke, $Y_{t+1-k}$, time of day, the interaction between $Y_{t+1-k}$ and time of day, baseline smoking severity, baseline drinking level, age and gender.


The data analysis leads to several conclusions.  First, the mindfulness message achieved a reduction in the average next-reported smoking rate, but only when the user was experiencing either a stable or decreased need to self-regulate ($95$\% CI $-5.45$ to $-0.15$ cigarettes per day; see \cref{tab:app}).  Otherwise no proximal treatment effect is apparent.  
Second, there is no evidence to support the presence of an overall lag-2 effect, with a $95$\% CI of $-1.74$ to $0.76$ cigarettes per day for the average reduction achieved by mindfulness treatment at the second-to-last treatment occasion.  Estimated standard errors (SEs) take into account sampling error in estimated treatment probabilities (see \cref{eq:varw} for the formula), and are corrected for small $n$ (see \cref{sec:implement} for details on the correction).
\begin{table}[H]
\centering
\caption{Proximal and lag-2 treatment effects estimated from BASICS-Mobile data.}
\label{tab:app}
\begin{tabular}{lcccc}
\toprule
Treatment effect & Estimate & SE & 95\% CI & $p$-value \\
\midrule
Proximal, increase in need to self-regulate    & $-0.06$ & 0.95 & $(-1.27,\phantom{-}1.16)$ & $0.99$ \\
Proximal, no increase in need to self-regulate & $-2.80$ & 1.29 & $(-5.45,-0.15)$ & $0.04$ \\
\midrule
Delayed                                        & $-0.49$ & 0.61 & $(-1.74,\phantom{-}0.76)$ & $0.43$ \\
\toprule
\end{tabular}
\end{table}

\section{Discussion}
\label{sec:discuss}

In this paper we define treatment effects suited for mobile interventions that enable frequent measurements and  frequent delivery of treatments.   As we discussed, the effect definition as provided in \cref{eq:prox} and \cref{eq:lagk} is atypical in the field of causal inference in that the underlying mechanism for the assigned treatment is part of the definition of the causal effect.
 However, this definition of the causal effects is consistent with the effects defined via most models for intensively collected longitudinal data (see \citealp{schafer2006}, \citealp{Stone2007} and, more recently, \citealp{bolger2013}).  Commonly the model for the conditional mean of a time-varying response given time-varying covariates is a linear model (possibly with the use of covariates defined by flexible basis functions).  If treatment indicators as well as interactions between the treatment indicators and time varying covariates are included in the linear model then the coefficients of these covariates coincide with the moderated proximal effect defined here.  However estimation of these casual effects using  most common approaches \citep{schafer2006, Stone2007, bolger2013}, that is, either GEE approaches or approaches that employ random effects, can cause bias.  Indeed the large sample and simulation results provided here show that straightforward use of GEEs (without weighting) is not guaranteed to consistently estimate $\beta_k^*$.

Since the conditional mean functions for models with random intercepts or random coefficients  \citep[e.g.][]{goldstein2011} are the same as those in GEEs, we expect that likelihood based methods which use the induced correlation structure in the estimation will generally be  biased.  This connection is important given the fact that, in the analysis of intensive longitudinal data, there is a preference for including random effects and, when GEE models are used, to use a non-independence working correlation structure (such as exchangeable, $\cor(Y_u, Y_t) = r$ ($u \neq t$), or $\text{AR}(1)$, $\cor(Y_u, Y_t) = r^{|u-t|}$) to improve precision \citep[p.~58]{schafer2006}. Indeed the large sample and simulation results provided here show that GEEs based on a non-independence working covariance structure is not guaranteed to consistently estimate $\beta_k^*$.  Future work is needed on whether or how to incorporate random effects in the estimation of proximal and lagged treatment effects.

There are a number of other directions for future work.  
Throughout we limited attention  to a continuous response and binary treatment decisions.   Lagged effects ($k>1$) were defined similar to proximal effects ($k=1$), but in future work one might rather be interested in a lagged effect that quantifies the accumulation of past treatment. Furthermore, since small to moderate treatment effects may be difficult to detect, yet potential response predictors that can be used in the working models to reduce error variance are numerous, future work could consider penalized methods for the working model in order to accommodate and select from the large number of predictors.  Lastly, here we considered analyses that are similiar to longitudinal analyses however interesting alternative approaches might have more of a ``system dynamics'' flavor and employ time-series modeling or Markovian process modeling. 

\bibliography{mHealthSNMMrefs}

\vspace{0.5in}
\noindent
{\it Acknowledgements:} Funding was provided by the National Institute on Drug Abuse (P50DA039838, R01DA039901, R01DA015697), National Institute on Alcohol Abuse and Alcoholism (R01AA023187), National Heart Lung and Blood Institute (R01HL125440), and National Institute of Biomedical Imaging and Bioengineering (U54EB020404).

\newpage
\appendix
\pagenumbering{arabic}
\renewcommand{\thepage}{A\arabic{page}}
\section*{Supplementary Material}

\section{Lagged Treatment Effects}
\label{app:snmm}

\subsection{Connection to Treatment Blips in the Structural Nested Mean Model}
\label{app:snmm:defn}

This \namecref{app:snmm} connects a generalization of the structural nested mean model (SNMM; \citealp{robins1989,robins1994}) to the lag $k$ treatment effect defined in \cref{sec:effects}. In particular, consider a causal effect or treatment ``blip'' as defined by the SNMM framework \citep[Section~3a]{robins1994}, with a minor departure in the choice of the reference treatment regime.  We show how these effects are additive on the conditional mean of the potential proximal response.  We conclude by connecting this particular SNMM generalization to the lag $k$ moderated effect \cref{eq:lagk} considered throughout the paper.

The typical reference treatment regime used to define the treatment ``blip'' functions under the SNMM framework, is  a prespecified non-random reference regime; here instead  our reference treatment regime is stochastic and will match the conditional distribution of the treatments given history in the data generating distribution.  In particular suppose that in the data generating distribution $\pr(A_t = 1 \mid \bar X_t = \bar x_t, \bar Y_t = \bar y_t,  \bar A_{t-1} = \bar a_{t-1}] =  p_t(1 \mid h_t)$ for each $t$ and where $h_t = (\bar x_t, \bar y_t,  \bar a_{t-1})$.  Then  the reference treatment regime for the potential  treatment is given, for each $t$, by $\pr(A_t(\bar a_{t-1}) = 1 \mid H_t(\bar a_{t-1}) = h_t] =  p_t(1 \mid h_t)$ (recall $H_t(\bar a_{t-1}) = (\bar X_t(\bar a_{t-1}), \bar Y_t(\bar a_{t-1}),  \bar A_{t-1}(\bar a_{t-2}))$).

The treatment blip of fixed $a_t\in\{0,1\}$ versus stochastic treatment $A_t(\bar a_{t-1})$ on the proximal response $Y_{t+1}$ is
\begin{equation*}
  \mu_{t,t+1}(h_t, \bar a_t)
  = \E\left[Y_{t+1}(\bar a_{t}) - Y_{t+1}(\bar a_{t-1},A_{t}(\bar a_{t-1}))
    \mid H_t(\bar a_{t-1}) = h_t\right].
\end{equation*}
  The treatment blip of fixed $a_{t-1}\in\{0,1\}$ versus stochastic treatment $A_{t-1}(\bar a_{t-2})$ on the response $Y_{t+1}$  is
\begin{multline*}
  \mu_{t-1,t+1}(h_{t-1}, \bar a_{t-1}) \\
  = \E\left[Y_{t+1}\left(\bar a_{t-1}, A_{t}(\bar a_{t-1})\right)
    - Y_{t+1}\left(\bar a_{t-2}, A_{t-1}(\bar a_{t-2}),
      A_{t}(\bar a_{t-2}, A_{t-1}(\bar a_{t-2}))\right)
    \mid H_{t-1}(\bar a_{t-2}) = h_{t-1}\right].
\end{multline*}
The treatment blip for general $u\le t$ is defined similarly but with an increase in notation.  However notice if we denote $A_2(A_1)$ by $A_2$ and so on with $A_t(\bar A_{t-1})$ denoted by $A_t$, and we denote $A_{u+1}(\bar A_{u-1}, a)$ by $A_{u+1}^{a_u=a}$, $A_{u+2}(\bar A_{u-1}, a, A_{u+1}^{a_u=a})$ by $A_{u+2}^{a_u=a}$ and so on with $A_t(\bar A_{u-1}, a, A_{u+1}^{a_u=a}, \ldots, A_{t-1}^{a_u=a})$ by $A_t^{a_u=a}$
 then we have the compact form
\begin{equation}\label{eq:blip}
  \mu_{u,t+1}(H_{u}(\bar A_{u-1}),\bar A_{u-1}, a)
  = \E\left[Y_{t+1}(\bar A_{u-1}, a_{u},A_{u+1}^{a_u=a},\ldots, A_t^{a_u=a})
  - Y_{t+1}(\bar A_{t}) \mid \bar H_{u}(\bar A_{u-1})\right].
\end{equation}

Assume consistency and sequential ignorability. 
 Then
\begin{align*}
  &\E\left[Y_{t+1}(\bar A_{u-1}, a, A_{u+1}^{a_u}, \ldots, A_t^{a_u = a})
    \bigmid H_{u}(\bar A_{u-1})\right] \\
  ={}& \E\left[Y_{t+1}(\bar A_{u-1}, a,A_{u+1}^{a_u = a}, \ldots, A_t^{a_u = a})
    \bigmid H_{u}(\bar A_{u-1}), A_u = a_u\right] \\
  ={}& \E\left[Y_{t+1}(\bar A_{u-1}, A_{u},A_{u+1}^{a_u = A_u},\ldots, A_t^{a_u = A_u})
    \bigmid H_{u}(\bar A_{u-1}), A_u = a_u\right] \\
  ={}& \E\left[Y_{t+1}(\bar A_t) \mid H_{u}(\bar A_{u-1}), A_u = a_u\right]
\end{align*}
where the first equality follows from the consistency and sequential ignorability assumptions (recall that $H_u=H_{u}(\bar A_{u-1})$) and the last two equalities follow by the definitions of $A_{j}^{a_u}$ and $A_j$.   Thus   the treatment blip  satisfies
\begin{equation}\label{eq:blip:zero}
  \E[\mu_{u,t+1}(H_u(\bar A_{u-1}), \bar A_{u})
  \mid H_u(\bar A_{u-1})] = 0,
\end{equation}
for each  $u = 1, \ldots, t$ and $t = 1, \ldots, T$. The lag $k$ treatment effect \cref{eq:lagk} can be expressed as the expected contrast of the treatment blips \cref{eq:blip}:
\begin{multline}\label{eq:lagk:snmm}
  \E\left[\mu_{t,t+k}(H_t(\bar A_{t-1}), \bar A_{t-1},  1)
    - \mu_{t,t+k}(H_t(\bar A_{t-1}), \bar A_{t-1}, 0)
    \mid S_{kt}(\bar A_{t-1})\right] \\
  = \E\left[Y_{t+k}(\bar A_{t-1}, 1, A_{t+1}^{a_u=1}, \ldots, A_{t+k-1}^{a_u=1})
    - Y_{t+k}(\bar A_{t-1}, 0, A_{t+1}^{a_u=0}, \ldots, A_{t+k-1}^{a_u=0})
    \bigmid S_{kt}(\bar A_{t-1})\right],
\end{multline}
given the candidate moderators $S_{kt}(A_{t-1})$.

As in \citet{robins1989,robins1994} the SNMM treatment blips are related to the conditional mean of $Y_{t+1}(\bar a_t)$ given $H_t(\bar a_{t-1})$ by way  of a telescoping sum.  For clarity we first provide the sum for $t=3$.
\begin{align*}
  & \E\left[Y_{4}(\bar a_3) \mid H_3(\bar a_{2}) = h_3\right] \\
  ={}& \E\left[Y_{4}(\bar a_3) - Y_{4}(\bar a_{2}, A_3(\bar a_2))
    \mid H_3(\bar a_{2}) = h_3\right] \\
  &+ \E\left[Y_{4}(\bar a_{2}, A_3(\bar a_2)) \mid H_3(\bar a_{2}) = h_3\right]
  - \E\left[Y_{4}(\bar a_{2}, A_3(\bar a_2))
    \mid H_{2}(\bar a_{1}) = h_{2}\right] \\
  &+ \E\left[Y_{4}(\bar a_{2}, A_3(\bar a_2))
    - Y_{4}({a}_{1}, A_2(a_1), A_3(a_1, A_2(a_1)))
    \mid H_{2}(\bar a_{1}) = h_{2}\right] \\
  &+ \E\left[Y_{4}({a}_{1}, A_2(a_1), A_3(a_1,A_2(a_1)))
    \mid H_{2}(\bar a_{1}) = h_{2}\right]
  - \E\left[Y_{4}({a}_{1}, A_2(a_1), A_3(a_1,A_2(a_1))) \mid H_{1} = h_{1}\right] \\
  &+ \E\left[Y_{4}({a}_{1}, A_2(a_1), A_3(a_1,A_2(a_1))) - Y_{4}(\bar A_3)
    \mid H_{1} = h_{1}\right] \\
  &+ \E\left[Y_{4}(\bar A_3) \mid H_{1} = h_{1}\right]
  - \E\left[Y_{4}(\bar A_3)\right] \\
  &+ \E\left[Y_{4}(\bar A_3)\right].
\end{align*}
Denote $A_{u+1}(\bar a_{u-1},A_u(\bar a_{u-1}))$ by $A_{u+1}^{\bar a_{u-1}}$, $A_{u+2}(\bar a_{u-1}, A_u(\bar a_{u-1}), A_{u+1}^{\bar a_{u-1}})$ by $A_{u+2}^{\bar a_{u-1}}$ and so on with $A_t\mathlist{(\bar a_{u-1},A_u(\bar a_{u-1}),A_{u+1}^{a_{u-1}}, \ldots, A_{t-1}^{\bar a_{u-1}})}$ by $A_t^{\bar a_{u-1}}$. Using this compact notation the treatment blips in \cref{eq:blip} can be rewritten as
\begin{equation*}
 \mu_{u,t+1}(h_u,\bar a_{u})
  = \E\left[Y_{t+1}(\bar a_{u}, A_{u+1}^{\bar a_u}, \ldots, A_t^{\bar a_u})
    - Y_{t+1}(\bar a_{u-1}, A_{u}^{\bar a_{u-1}}, \ldots, A_t^{\bar a_{u-1}})
    \mid \bar H_{u}(\bar a_{u-1}) = h_u\right].
\end{equation*}
The telescoping sum for general $t$ using this compact notation is
\begin{align}\label{eq:telsum}
  & \E[Y_{t+1}(\bar a_t) \mid H_t(\bar a_{t-1}) = h_t] \notag \\
  ={}& \E\left[Y_{t+1}(\bar a_t) - Y_{t+1}(\bar a_{t-1}, A_t^{\bar a_{t-1}})
    \mid H_t(\bar a_{t-1}) = h_t\right] \notag \\
  &+ \E[Y_{t+1}(\bar a_{t-1}, A_t^{\bar a_{t-1}}) \mid H_t(\bar a_{t-1}) = h_t]
  - \E[Y_{t+1}(\bar a_{t-1}, A_t^{\bar a_{t-1}})
  \mid H_{t-1}(\bar a_{t-2})=h_{t-1}] \notag \\
  &+ \E[Y_{t+1}(\bar a_{t-1}, A_t^{\bar a_{t-1}})
  - Y_{t+1}(\bar a_{t-2}, A_{t-1}^{\bar a_{t-2}},A_{t}^{\bar a_{t-2}})
  \mid H_{t-1}(\bar a_{t-2})=h_{t-1}] \notag \\
  &+ \E[Y_{t+1}(\bar a_{t-2}, A_{t-1}^{\bar a_{t-2}}, A_{t}^{\bar a_{t-2}})
  \mid H_{t-1}(\bar a_{t-2}) = h_{t-1}]
  - \E[Y_{t+1}(\bar a_{t-2}, A_{t-1}^{\bar a_{t-2}}, A_{t}^{\bar a_{t-2}})
  \mid H_{t-2}(\bar a_{t-3}) = h_{t-2}] \notag \\
  &\cdots \notag \\
  &+ \E[Y_{t+1}({a}_{1}, A_2^{a_1},\cdots, A_t^{a_1})
  - Y_{t+1}(\bar A_t) \mid H_{1} = h_{1}] \notag \\
  &+ \E[Y_{t+1}(\bar A_t)\mid H_{1} = h_{1}]- \E[Y_{t+1}(\bar A_t)] \notag \\
  &+ \E[Y_{t+1}(\bar A_t)] \notag \\
  ={}& \E[Y_{t+1}(\bar A_t)] + \sum_{u=1}^t \mu_{u,t+1}(h_u, \bar a_u)
  + \sum_{u=1}^t \epsilon_{u,t+1}(h_u, \bar a_{u-1}),
\end{align}
where
\begin{align*}
  \epsilon_{u,t+1}(h_u, \bar a_{u-1})
  ={}& \E[Y_{t+1}(\bar a_{u-1}, A_u^{\bar a_{u-1}}, \ldots, A_t^{\bar a_{u-1}})
  \mid H_u(\bar a_{u-1}) = h_u] \\
  &- \E[Y_{t+1}(\bar a_{u-1}, A_u^{\bar a_{u-1}}, \ldots, A_t^{\bar a_{u-1}})
  \mid H_{u-1}(\bar a_{u-2}) = h_{u-1}],
\end{align*}
are nuisance functions that satisfy the constraint $\E[\epsilon_{u,t+1}(H_u(\bar a_{u-1}), \bar a_{u-1}) \mid H_{u-1}(\bar a_{u-2})] = 0$, for each $\bar a_{u-1} \in \mathcal{A}_{u-1}$, $u = 1, \ldots, t$ and $t = 1, \ldots, T$.

\subsection{Identification from Data}
\label{app:snmm:obs}

Here we derive the expression \cref{eq:lagk:obs} of the lag $k$ treatment effect \cref{eq:lagk}.  This is done under the consistency, positivity and sequential ignorability conditions described in \cref{sec:effects}.

To derive expression \cref{eq:lagk:obs} for the lag $k$ treatment effect \cref{eq:lagk}, we show that
\begin{equation*}
  \E\left[Y_{t+k}(\bar A_{t-1}, a, A_{t+1}^{a_t=a}, \ldots, A_{t+k-1}^{a_t=a})
  \bigmid S_{kt}(\bar A_{t-1})\right]
  = \E\left[\E[Y_{t+k} \mid A_t = a, H_t] \bigmid S_{kt}\right]
\end{equation*}
and
\begin{equation*}
  \E\left[Y_{t+k}(\bar A_{t-1}, a, A_{t+1}^{a_t=a}, \ldots, A_{t+k-1}^{a_t=a})
   \mid S_{kt}(\bar A_{t-1})\right]
  = \E\left[\frac{\indic(A_t = a)}{ p_t(a \mid H_t)}
    Y_{t+k} \Bigmid S_{kt}\right]
\end{equation*}
for $a\in \{0,1\}$.

First recall that by consistency, $H_t=H_t(\bar A_{t-1})$ and $S_{kt}=S_{kt}(\bar A_{t-1})$.   Second recall the definition of $A_{t+j}^{a_t = a}$, where in particular $A_{t+1}^{a_t = a}$ denotes $A_{t+1}(\bar A_{t-1}, a)$, $A_{t+2}^{a_t=a}$ denotes $A_{t+2}(\bar A_{t-1}, a, A_{t+1}^{a_t=a})$ and so on, with $A_{t+k-1}\mathlist{(\bar A_{t-1},a, A_{t+1}^{a_t}, \ldots, A_{t+k-2}^{a_t=a})}$ denoted by $A_{t+k-1}^{a_t=a}$).   So for each $j=1, \ldots, T-t+1$,  sequential ignorability implies that $A_{t+j}^{a_t=a}, a\in\{0,1\} $ is independent of $A_t$ given $H_t$.
We have
\begin{align*}
  & \E[Y_{t+k}(\bar A_{t-1}, a, A_{t+1}^{a_t=a}, \ldots, A_{t+k-1}^{a_t=a})
  \mid S_{kt}(\bar A_{t-1})] \\
  ={}& \E\left[\E\left[Y_{t+k}(\bar A_{t-1}, a, A_{t+1}^{a_t=a}, \ldots, A_{t+k-1}^{a_t=a})
      \mid H_{t}(\bar A_{t-1})\right] \bigmid S_{kt}(\bar A_{t-1})\right] \\
  ={}& \E\left[\E\left[Y_{t+k}(\bar A_{t-1}, a, A_{t+1}^{a_t=a}, \ldots, A_{t+k-1}^{a_t=a})
      \mid H_{t}\right] \bigmid S_{kt}\right] \\
  ={}& \E\left[\E\left[Y_{t+k}(\bar A_{t-1}, a, A_{t+1}^{a_t=a}, \ldots, A_{t+k-1}^{a_t=a})
      \mid H_{t}, A_t=a\right] \bigmid S_{kt}\right] \\
  ={}& \E\left[\E\left[Y_{t+k}(\bar A_{t-1}, A_t, A_{t+1}^{a_t=A_t},
      \ldots, A_{t+k-1}^{a_t=A_t})
      \mid H_{t}, A_t=a\right] \bigmid S_{kt}\right] \\
        ={}& \E\left[\E\left[Y_{t+k}(\bar A_{t-1}, A_t, A_{t+1}, \ldots, A_{t+k-1})
      \mid H_{t}, A_t=a\right] \bigmid S_{kt}\right] \\
  ={}& \E\left[\E\left[Y_{t+k} \mid H_{t}, A_t=a\right] \bigmid S_{kt}\right],
\end{align*}
where the second equality holds by consistency, the third by sequential ignorability and the fifth follows from the  definition of $A_{t+j}^{a_t=a}$ implying that $A_{t+j}^{a_t=A_t}=A_{t+j}$.

Next note that, by sequential ignorability, $\E[Y_{t+k}(\bar A_{t-1}, a, A_{t+1}^{a_t=a}, \ldots, A_{t+k-1}^{a_t=a})\mid H_{t}] \E[\indic(A_t = a)\mid H_t]$ is equal to $\E[Y_{t+k}(\bar A_{t-1}, a, A_{t+1}^{a_t = a}, \ldots, A_{t+k-1}^{a_t = a}) \indic(A_t = a) \mid H_{t}]$.  We have
\begin{align*}
  & \E\left[Y_{t+k}(\bar A_{t-1}, a, A_{t+1}^{a_t=a}, \ldots, A_{t+k-1}^{a_t=a})
  \bigmid S_{kt}(\bar A_{t-1})\right] \\
  ={}& \E\left[\E\left[Y_{t+k}(\bar A_{t-1}, a, A_{t+1}^{a_t=a}, \ldots, A_{t+k-1}^{a_t=a})
      \bigmid H_{t}\right] \Bigmid S_{kt}\right] \\
  ={}& \E\left[\E\left[Y_{t+k}(\bar A_{t-1}, a, A_{t+1}^{a_t=a}, \ldots, A_{t+k-1}^{a_t=a})
      \bigmid H_{t}\right] \frac{\E[\indic(A_t = a)\mid H_t]}{ p_t(a \mid H_t)}
    \Bigmid S_{kt}\right] \\
  ={}& \E\left[\E\left[Y_{t+k}(\bar A_{t-1}, a, A_{t+1}^{a_t=A_t},
      \ldots, A_{t+k-1}^{a_t=A_t}) \frac{\indic(A_t = a)}{ p_t(a \mid H_t)}
      \Bigmid H_{t}\right] \Bigmid S_{kt}\right] \\
  ={}& \E\left[\E\left[Y_{t+k}\frac{\indic(A_t = a)}{ p_t(a \mid H_t)}
      \Bigmid H_{t}\right] \Bigmid S_{kt}\right] \\
  ={}& \E\left[Y_{t+k}\frac{\indic(A_t = a)}{ p_t(a \mid H_t)}
    \Bigmid S_{kt}\right]
\end{align*}

\section{Model Specification}
\label{app:model}

This \lcnamecref{app:model} discusses why the treatment effect at a given lag can be modeled without consideration of treatment effect models  at other lags.  We also provide a simple example of how models for   $\E[W_tY_{t+k} \mid H_t]$ at different lags $k$ constrain one another and are constrained by and constrain the treatment effect models.  These considerations lead us to avoid assumptions concerning the correctness of  models for $\E[W_tY_{t+k} \mid H_t]$. 
 For clarity we consider the case in which $W_t=1$ for all $t$ and thus illustrate why we avoid assumptions concerning the correctness of models for $\E[Y_{t+k} \mid H_t]$.  

From \cref{eq:lagk:snmm}, we know that the lag $k$ effect depends on only one of the SNMM treatment blips \cref{eq:blip}.  From \cref{eq:telsum} these blips are in turn additive on the conditional mean of the potential response.  Provided that this conditional mean is not \emph{a priori} restricted to certain values in $(-\infty, \infty)$, the treatment blips do not constrain one another \citep[Theorem~8.6]{robins2000c}.  This implies the same result for the lag $k$ effect; that is, the treatment effects at different lags can be specified separately, with each lag-specific model imposing no constraints on the models chosen for the treatment effects at the remaining lags.

As an example, here we provide an illustration of how a model chosen for the lag $1$ conditional mean response $\E[Y_{t+1} \mid H_t]$ constrains  the form of the treatment effects at lag 2.  Consider the simple example in which the treatments are binary, randomized with probability $0.5$.   Suppose we model the conditional mean of the response, $\E[Y_{t+1} \mid  H_t]$  by
$\alpha_{10} + \alpha_{11} Z_t+ \alpha_{12}A_{t-1}$, where $Z_t$ is an  binary variable influenced by $A_{t-1}$.  Further suppose that we model the  lag 2 treatment effect, $\E[Y_{t+1} \mid A_{t-1} = 1, H_{t-1}]
  - \E[Y_{t+1} \mid A_{t-1} = 0, H_{t-1}]$ by a linear model $H_{t-1}^\tr\beta_2$.  Unfortunately in general these two models are inconsistent; they cannot both be correct.   To see this, suppose that unbeknownst to us, $\pr[Z_t=1 \mid H_{t-1}] = 1/(1 + \exp(Y_{t-1} + A_{t-1}))$.  Now if the first model is correct then the true lag-2 treatment effect should satisfy
\begin{align*}
  &\E[Y_{t+1} \mid A_{t-1} = 1, H_{t-1}]
  - \E[Y_{t+1} \mid A_{t-1} = 0, H_{t-1}] \\
  ={}& \E[\E[Y_{t+1} \mid H_t] \mid A_{t-1} = 1, H_{t-1}]
  - \E[\E[Y_{t+1} \mid  H_t] \mid A_{t-1} = 0, H_{t-1}] \\
  ={}& \alpha_{11}  \left\{\pr[Z_t = 1 \mid A_{t-1} = 1, H_{t-1}]
  - \pr(Z_t = 1 \mid A_{t-1} = 0, H_{t-1})\right\} + \alpha_{12}\\
  ={}& \alpha_{11}  \left\{\frac{1}{1+e^{Y_{t-1}+1}}-\frac{1}{1+e^{Y_{t-1}}}\right\}+ \alpha_{12}.
\end{align*}
In general since the conditional probability of $Z_t = 1$ is constrained to $[0, 1]$, this expression will be non-linear in $H_{t-1}$.  So these lag 2 treatment effect and the lag 1 conditional mean response models cannot both be true.

This example shows that both parsimony in the treatment effect models  and correctness in the models for the conditional mean response is difficult to achieve in the presence of binary (or more generally non-continuous) response predictors.  Two special scenarios in which models with main effect of the form $g_{kt}(H_t)^\tr\alpha_k$ might be coherent across different $k$ arise when all variables in $g_{kt}(H_t)$ are either (1) multivariate normal, or (2) centered by their conditional mean---i.e., $g_{kt}(H_t)$ is replaced by $g_{kt}(H_t) - \E[g_{kt}(H_t) \mid H_{t-1}]$---since $E[ g_{kt}(H_t) - \E[g_{kt}(H_t) \mid H_{t-1}] ] = 0$.  Both of these settings require strong restrictions or additional assumptions about the distribution of covariates.  So in general we should prefer estimation methods where $g_{kt}(H_t)^\tr\alpha_k$ need only be a working model for $\E[W_tY_{t+k} \mid  H_t]$.

\section{Large Sample Properties}
\label{app:large}

In this \lcnamecref{app:large} we derive the large sample properties stated in \cref{sec:estimate}. 
 Throughout we allow for the setting in which individuals are not always available as discussed in \cref{sec:avail}. For completeness we provide results for a more general estimating function which can be used with observational (non-randomized $A_t$) treatments, under the assumption of sequential ignorability and assuming the data analyst is able to correctly model and estimate the treatment probability, $P[A_t=1\mid H_t]$.  We indicate how the results are simplified by use of data from an MRT.

 Denote the parameterized treatment probability by $p_t(1\mid H_t;\eta)$ (with parameter $\eta$); note $\eta$ is known in an MRT.  Denote the parameterized numerator of the weights by $\tilde p_t(1\mid S_{kt};\rho)$ (with parameter $\rho$); below in (\ref{eq:projbeta}) we will see that the numerator of the weights defines the estimand for $\hat\beta_k$ when our modeling assumption (\ref{eq:withavail}) is incorrect. In this case, the estimator  $\hat\beta_k$ converges to the weights on a projection defined by $\tilde p_t$.  The proof below allows the data analyst  to use a  $\tilde p_t$ with an estimated parameter, $\hat\rho$ or to pre-specify $\rho$ as desired.  
 We use a superscript of $*$ to denote limiting values of estimated parameters (e.g. $\eta^*, \rho^*$).  
 Then the more general version of the estimating equation \cref{eq:lsw} is  
\begin{multline}\label{eq:geew}
 U_{\mathrm{W}}(\alpha_k, \beta_k;  \hat\eta, \hat\rho)\\
 =\sum_{t=1}^{T-k+1} \left(Y_{t+k} - g_{kt}(H_t)^\tr\alpha_k
    - (A_t-\tilde p_t(1|S_{kt};\hat\rho)) f_{kt}(S_{kt})^\tr\beta_k\right) I_t W_t(A_t, H_t; \hat\eta, \hat\rho)  \\
  \begin{pmatrix} g_{kt}(H_t) \\ (A_t-\tilde p_t(1|S_{kt};\hat\rho))  f_{kt}(S_{kt}) \end{pmatrix}
\end{multline}
where $W_t(A_t, H_t; \eta, \rho) = \tilde p_t(A_t\mid S_{kt};\rho)/ p_t(A_t \mid H_t; \eta)$ and $\hat\eta,\hat\rho$ are estimators.  Note $W_t$ in the body of the paper is replaced here by $W_t(A_t, H_t; \hat\eta, \hat\rho)$.

 Throughout we assume the model, \cref{eq:withavail},  and sequential ignorability.   Assume the following for the $k$ lags of interest.
\begin{assumew}
\item\label{wt:moment}  All entries in $\{Y_{t+k}, g_{kt}(H_t)\}_{t=1}^{t=T-k+1}$ have finite fourth moments.
\item\label{wt:invert}  The matrices  $\E\left[\sum_t  I_t S_{kt}^{\otimes 2} \right]$ and 
$$\E\dot U_{\mathrm{W}}( \eta^*, \rho^*)=  E\sum_t \sum_a  I_t \tilde p_t(a| S_{kt}; \rho^*)
  \begin{pmatrix} g_{kt}(H_t) \\ (a-\tilde p_t(1|S_{kt};\rho^*))  f_{kt}(S_{kt}) \end{pmatrix}^{\otimes 2}$$
are invertible.
\end{assumew}
 If the data is observational then we  assume:
  \begin{assume}
  \item\label{con:prob} Treatment Probability Model:  $ p_t(1 \mid H_t; \eta)$ is a correctly specified  model for $\pr(A_t = 1 \mid I_t = 1, H_t)$.  Let $\eta^*$ be the true value of $\eta$; that is, $\pr(A_t = 1 \mid I_t = 1, H_t) =  p_t(1 \mid H_t; \eta^*)$.  Assume that the estimator of $\eta$, say $\hat\eta$, satisfies $\eprob U_{\mathrm{D}}(\hat\eta) = 0$ and $\sqrt{n}(\hat\eta - \eta^*) = \E[\dot U_{\mathrm{D}}(\eta^*)]^{-1}\eprob U_{\mathrm{D}}(\eta^*) + o_P(1)$.  Thus $\sqrt{n}(\hat\eta - \eta^*)$ converges in distribution to a mean zero, Normal random vector with variance-covariance matrix given by $\E[\dot U_{\mathrm{D}}(\eta^*)]^{-1}\E[U_{\mathrm{D}}(\eta^*)^{\otimes 2}] (\E[\dot U_{\mathrm{D}}(\eta^*)]^{-1})^\tr$ which has finite entries.   Assume that  $\eprob \dot U_{\mathrm{D}}(\hat\eta)$ is a consistent estimator of $\E[\dot U_{\mathrm{D}}(\eta^*)]$.  Assume there exists finite constants, $b_{\mathrm{D}} > 0$ and $B_{\mathrm{D}} < 1$ such that each $b_{\mathrm{D}} <  p_t(1 \mid H_t; \eta^*) < B_{\mathrm{D}}$ a.s.
  \end{assume}
   If the data analyst elects to use a parameterized and estimated $\tilde p_t(1|S_{kt},\hat\rho)$, then we assume:
\begin{assume}
\item\label{con:stable} Numerator of Weights Probability Model:  Suppose the estimator $\hat\rho$ solves an estimating equation: $\eprob U_{\mathrm{N}}(\rho)=0$.  Assume that, for  a finite value of $\rho$, say $\rho^*$ and $\sqrt{n}(\hat\rho - \rho^*) = \E[\dot U_{\mathrm{N}}(\rho^*)]^{-1} \sqrt{n}(\eprob - \prob) U_{\mathrm{N}}(\rho^*) + o_P(1)$ where the matrix, $\E[\dot U_{\mathrm{N}}(\rho^*)]$ is positive definite.  Assume $\sqrt{n}(\eprob - \prob) U_{\mathrm{N}}(\rho^*)$ converges in distribution to a mean zero, Normal random vector with variance-covariance matrix given by $\E[U_{\mathrm{N}}(\rho^*)^{\otimes 2}]$ which has finite entries.  Assume that  $\eprob \dot U_{\mathrm{N}}(\hat\rho) $ is a consistent estimator of $\E[\dot U_{\mathrm{N}}(\rho^*)]$.  Assume $0 < \rho^* < 1$.
\end{assume}



The solution to $\eprob U_{\mathrm{W}}(\alpha_k, \beta_k; \hat\eta,\hat\rho) = 0$ gives the estimator
\begin{equation*}
  \begin{pmatrix} \hat{\alpha}_k \\ \hat{\beta}_k \end{pmatrix}
  = \left\{\eprob \dot U_{\mathrm{W}}( \hat\eta, \hat\rho)\right\}^{-1}
  \eprob \sum_t
  I_t W_t(A_t, H_{t}; \hat\eta, \hat\rho) Y_{t+k}
  \begin{pmatrix} g_{kt}(H_t) \\ (A_t-\tilde p_t(1|S_{kt};\hat\rho))  f_{kt}(S_{kt})\end{pmatrix}
\end{equation*}
where
\begin{equation*}
   \dot U_{\mathrm{W}}(\eta, \rho)
  =  \sum_t 
  I_t W_t(A_t, H_t; \eta, \rho)
  \begin{pmatrix} g_{kt}(H_t) \\ (A_t-\tilde p_t(1|S_{kt};\rho))  f_{kt}(S_{kt}) \end{pmatrix}^{\otimes 2}.
\end{equation*}

Define
\begin{equation*}
  \begin{pmatrix} \alpha_k' \\ \beta_k' \end{pmatrix}
  = \left\{\E\left[\dot U_{\mathrm{W}}( \eta^*, \rho^*)\right]\right\}^{-1}
  \E\left[\sum_t 
    I_t W_t(A_t, H_{t}; \eta^*, \rho^*) Y_{t+k}
    \begin{pmatrix} g_{kt}(H_t) \\ (A_t-\tilde p_t(1|S_{kt};\rho^*))  f_{kt}(S_{kt}) \end{pmatrix}\right].
\end{equation*}
Then standard statistical arguments can be used to show that $\sqrt{n}(\hat\alpha_k - \alpha_k', \hat\beta_k - \beta_k')$ converges in distribution to a normal, mean zero, random vector with variance-covariance matrix given by
\begin{equation*}
  \left\{\E\left[\dot U_{\mathrm{W}}(\eta^*, \rho^*)\right]\right\}^{-1}
  \Sigma_{\mathrm{W}}(\alpha_k', \beta_k';  \eta^*, \rho^*)
  \left\{\E\left[\dot U_{\mathrm{W}}(\eta^*, \rho^*)\right]\right\}^{-1},
 \end{equation*}
where
\begin{align*}
  \Sigma_{\mathrm{W}}(\alpha_k, \beta_k;   \eta, \rho)
  = \E\bigg[\bigg({U}_{\mathrm{W}}(\alpha_k, \beta_k;   \eta, \rho)
  &+ \Sigma_{\mathrm{W,D}}(\alpha_k, \beta_k;   \eta, \rho)
  \{\E[\dot U_{\mathrm{D}}(\eta)]\}^{-1} U_{\mathrm{D}}(\eta) \\
  &+ \Sigma_{\mathrm{W,N}}(\alpha_k, \beta_k;   \eta, \rho)
  \{\E[\dot U_{\mathrm{N}}(\rho)]\}^{-1} U_{\mathrm{N}}(\rho)\bigg)^{\otimes 2}\bigg],
\end{align*}
with
\begin{multline*}
  \Sigma_{\mathrm{W,D}}(\alpha_k, \beta_k;   \eta, \rho) \\
  =\E\bigg[\sum_{t=1}^{T-k+1} \left(Y_{t+k}
      - g_{kt}(H_t)^\tr\alpha_k -(A_t-\tilde p_t(1|S_{kt};\rho))  f_{kt}(S_{kt}) ^\tr\beta_k\right)
    I_t  W_t(A_t, H_t; \eta, \rho)\\
    \begin{pmatrix} g_{kt}(H_t) \\ (A_t-\tilde p_t(1|S_{kt};\rho))  f_{kt}(S_{kt})  \end{pmatrix}
    \left(\frac{d \log p_t(A_t \mid H_t; \eta)}{d\eta}\right)^\tr\bigg],
\end{multline*}
 and  
\begin{multline*}
 \Sigma_{\mathrm{W,N}}(\alpha_k, \beta_k;   \eta, \rho)\\
 = \E\bigg[\sum_{t=1}^{T-k+1} \left(Y_{t+k}
      - g_{kt}(H_t)^\tr\alpha_k - (A_t-\tilde p_t(1|S_{kt};\rho))  f_{kt}(S_{kt}) ^\tr\beta_k\right)
    I_t W_t(A_t, H_t; \eta, \rho) \\
   \phantom{bbbbbbbbbbbbbbbbbbbbbbbb} \begin{pmatrix} g_{kt}(H_t) \\ (A_t-\tilde p_t(1|S_{kt};\rho))  f_{kt}(S_{kt})  \end{pmatrix}
    \left( \frac{d \log \tilde p_t(A_t \mid S_{kt}; \rho)}{d\rho}\right)^\tr\bigg]\\
    +  \E\bigg[\sum_{t=1}^{T-k+1} \left(Y_{t+k}
      - g_{kt}(H_t)^\tr\alpha_k - (A_t-\tilde p_t(1|S_{kt};\rho))  f_{kt}(S_{kt}) ^\tr\beta_k\right)
    I_t W_t(A_t, H_t; \eta, \rho) \\
    \phantom{bbbbbbbbbbbbbbbbbbbb} \begin{pmatrix} 0_{q\times 1} \\ -\tilde p_t(1|S_{kt};\rho)f_{kt}(S_{kt}) \end{pmatrix}  \left(\frac{d \log \tilde p_t(1 \mid S_{kt}; \rho)}{d\rho}\right)^\tr\bigg]\\
    + \E\bigg[\sum_{t=1}^{T-k+1} \tilde p_t(1|S_{kt};\rho)f_{kt}(S_{kt}) ^\tr\beta_k    I_t W_t(A_t, H_t; \eta, \rho)
    \begin{pmatrix} g_{kt}(H_t) \\ (A_t-\tilde p_t(1|S_{kt};\rho))  f_{kt}(S_{kt})  \end{pmatrix}\\
   \phantom{bbbbbbbbb}  \left(\frac{d \log \tilde p_t(1 \mid S_{kt}; \rho)}{d\rho}\right)^\tr\bigg]
\end{multline*}
where $q$ is the dimension of $\alpha_k$.
Note that if the data is from a MRT (we know $p_t$) and we pre-specify (not estimate) $\tilde p_t$ then $ \Sigma_{\mathrm{W}}(\alpha_k, \beta_k;   \eta, \rho)
  = \E\bigg[\bigg({U}_{\mathrm{W}}(\alpha_k, \beta_k;   \eta, \rho)
  \bigg)^{\otimes 2}\bigg]$  greatly simplifying the variance-covariance matrix.

A consistent estimator of the variance-covariance matrix is given by
\begin{equation}\label{eq:varw}
  \left\{\eprob\dot U_{\mathrm{W}}( \hat\eta, \hat\rho)\right\}^{-1}
  \hat\Sigma_{\mathrm{W}}(\hat\alpha_k, \hat\beta_k; \hat\eta, \hat\rho)
  \left\{\eprob\dot U_{\mathrm{W}}(\hat\eta, \hat\rho)\right\}^{-1},
\end{equation}
where
\begin{multline*}
  \hat\Sigma_{\mathrm{W}}(\alpha_k, \beta_k;\eta,\rho)
  = \eprob \bigg[\bigg({U}_{\mathrm{W}}(\alpha_k, \beta_k; \eta, \rho) 
  + \hat\Sigma_{\mathrm{W,D}}(\alpha_k, \beta_k; \eta, \rho)
  \{\eprob\dot U_{\mathrm{D}}(\eta)\}^{-1} U_{\mathrm{D}}(\eta)\\
 + \hat\Sigma_{\mathrm{W,N}}(\alpha_k, \beta_k;  \eta, \rho)
  \{\eprob\dot U_{\mathrm{N}}(\rho)\}^{-1} U_{\mathrm{N}}(\rho)\bigg)^{\otimes 2}\bigg],
\end{multline*}
with
\begin{multline*}
  \hat\Sigma_{\mathrm{W,D}}(\alpha_k, \beta_k; \gamma, \eta, \rho) \\
  = \eprob\bigg[\sum_{t=1}^{T-k+1} \left(Y_{t+k}
      - g_{kt}(H_t)^\tr\alpha_k -(A_t-\tilde p_t(1|S_{kt};\rho))  f_{kt}(S_{kt}) ^\tr\beta_k\right)
    I_t  W_t(A_t, H_t; \eta, \rho)\\
    \begin{pmatrix} g_{kt}(H_t) \\ (A_t-\tilde p_t(1|S_{kt};\rho))  f_{kt}(S_{kt})  \end{pmatrix}
    \left(\frac{d \log p_t(A_t \mid H_t; \eta)}{d\eta}\right)^\tr\bigg]
\end{multline*}
and  $  \hat\Sigma_{\mathrm{W,N}}(\alpha_k, \beta_k; \gamma, \eta, \rho)
  = $
\begin{multline*}
\eprob\bigg[\sum_{t=1}^{T-k+1} \left(Y_{t+k}
      - g_{kt}(H_t)^\tr\alpha_k - (A_t-\tilde p_t(1|S_{kt};\rho))  f_{kt}(S_{kt}) ^\tr\beta_k\right)I_t W_t(A_t, H_t; \eta, \rho)  \\
\phantom{bbbbbbbbbbbbbbbbbbbbbbbb} \begin{pmatrix} g_{kt}(H_t) \\ (A_t-\tilde p_t(1|S_{kt};\rho))  f_{kt}(S_{kt})  \end{pmatrix}
    \left( \frac{d \log \tilde p_t(A_t \mid S_{kt}; \rho)}{d\rho}\right)^\tr\bigg]\\
    +  \eprob\bigg[\sum_{t=1}^{T-k+1} \left(Y_{t+k}
      - g_{kt}(H_t)^\tr\alpha_k - (A_t-\tilde p_t(1|S_{kt};\rho))  f_{kt}(S_{kt}) ^\tr\beta_k\right)
    I_t W_t(A_t, H_t; \eta, \rho) \\
\phantom{bbbbbbbbbbbbbbbbbbbbbbbbbbbbb}     \begin{pmatrix} 0_{q\times 1} \\ -\tilde p_t(1|S_{kt};\rho)f_{kt}(S_{kt}) \end{pmatrix}\left( \frac{d \log \tilde p_t(1 \mid S_{kt}; \rho)}{d\rho}\right)^\tr\bigg]\\
    +\eprob\bigg[\sum_{t=1}^{T-k+1} \tilde p_t(1|S_{kt};\rho)f_{kt}(S_{kt}) ^\tr\beta_k    I_t W_t(A_t, H_t; \eta, \rho) 
     \begin{pmatrix} g_{kt}(H_t) \\ (A_t-\tilde p_t(1|S_{kt};\rho))  f_{kt}(S_{kt})  \end{pmatrix}\\
     \left( \frac{d \log \tilde p_t(1 \mid S_{kt}; \rho)}{d\rho}\right)^\tr\bigg].
\end{multline*}

It remains to show that $\beta_k'=\beta_k^*$. Since $\E[U_{\mathrm{W}}(\alpha_k', \beta_k'; \gamma^*, \eta^*, \rho^*)] = 0$,
\begin{eqnarray*}
  0 &=& \E\sum_{t=1}^{T-k+1} \left(Y_{t+k}
    - g_{kt}(H_t)^\tr\alpha_k' -  (A_t-\tilde p_t(1|S_{kt};\rho^*)) f_{kt}(S_{kt}) ^\tr\beta_k'\right)\\
& & \phantom{bbbbbbbbbbbbbbbbbbbbbb} I_t w_{t}(A_t,H_{t};\eta^*,\rho^*)
  (A_t-\tilde p_t(1|S_{kt};\rho^*)) f_{kt}(S_{kt}) \\
  & =&\E \sum_{t=1}^{T-k+1}  \left(\E\left[Y_{t+k}\mid A_t, H_t, I_t = 1\right]
    - g_{kt}(H_t)^\tr\alpha_k' -  (A_t-\tilde p_t(1|S_{kt};\rho^*)) f_{kt}(S_{kt}) ^\tr\beta_k'\right)\\
& & \phantom{bbbbbbbbbbbbbbbbbbbbb}  I_t w_{t}(A_t, H_{t}; \eta^*, \rho^*)
   (A_t-\tilde p_t(1|S_{kt};\rho^*)) f_{kt}(S_{kt})   \\
  &=&\E\sum_{t=1}^{T-k+1}\sum_{a \in \{0,1\}}
  \left(\E\left[Y_{t+k}\mid A_t = a, H_t, I_t = 1\right]
    - g_{kt}(H_t)^\tr\alpha_k' -  (a-\tilde p_t(1|S_{kt};\rho^*)) f_{kt}(S_{kt}) ^\tr\beta_k'\right)\\
 & &   \phantom{bbbbbbbbbbbbbbbbbbbbb}I_t \tilde p_t(a|S_{kt};\rho^*)
  (a-\tilde p_t(1|S_{kt};\rho^*)) f_{kt}(S_{kt})  
\end{eqnarray*}
where the last equality averages out over $A_t$.   The above simplifies to,
\begin{eqnarray*}
  0   &=&\E\sum_{t=1}^{T-k+1}\sum_{a \in \{0,1\}}
  \left(\E\left[Y_{t+k}\mid A_t = a, H_t, I_t = 1\right]
    - g_{kt}(H_t)^\tr\alpha_k' -  (a-\tilde p_t(1|S_{kt};\rho^*)) f_{kt}(S_{kt}) ^\tr\beta_k'\right)\\
 & &   \phantom{bbbbbbbbbbbbbbbbbbbbb}I_t \tilde p_t(a|S_{kt};\rho^*)
  (a-\tilde p_t(1|S_{kt};\rho^*)) f_{kt}(S_{kt}) \\ 
  &=& \E\sum_{t=1}^{T-k+1}  \left(\E\left[Y_{t+k}\mid A_t = 1, H_t, I_t = 1\right]
    - g_{kt}(H_t)^\tr\alpha_k' -  (1-\tilde p_t(1|S_{kt};\rho^*)) f_{kt}(S_{kt}) ^\tr\beta_k'\right)\\
 & &   \phantom{bbbbbbbbbbbbbbbbbbbbb}I_t \tilde p_t(1|S_{kt};\rho^*)
  (1-\tilde p_t(1|S_{kt};\rho^*)) f_{kt}(S_{kt})\\
  & & + \left(\E\left[Y_{t+k}\mid A_t = 0, H_t, I_t = 1\right]
    - g_{kt}(H_t)^\tr\alpha_k' -  (-\tilde p_t(1|S_{kt};\rho^*)) f_{kt}(S_{kt}) ^\tr\beta_k'\right)\\
 & &   \phantom{bbbbbbbbbbbbbbbbbbbbb}I_t (1-\tilde p_t(1|S_{kt};\rho^*))
  (-\tilde p_t(1|S_{kt};\rho^*)) f_{kt}(S_{kt})\\
   &=& \E\sum_{t=1}^{T-k+1} f_{kt}(S_{kt}) (1-\tilde p_t(1|S_{kt};\rho^*))\tilde p_t(1|S_{kt};\rho^*)I_t\\
& &\phantom{bbbbbbbbb}   \big(\E\left[Y_{t+k}\mid A_t = 1, H_t, I_t = 1\right]-\E\left[Y_{t+k}\mid A_t = 0, H_t, I_t = 1\right]     -   f_{kt}(S_{kt}) ^\tr\beta_k'\big).
\end{eqnarray*}
From this we obtain,
\begin{eqnarray*}
 0  &=& \E\sum_{t=1}^{T-k+1} f_{kt}(S_{kt}) (1-\tilde p_t(1|S_{kt};\rho^*))\tilde p_t(1|S_{kt};\rho^*)I_t\\
& &\phantom{bbbbbbbbb}   \Big(\E\big[\E\left[Y_{t+k}\mid A_t = 1, H_t, I_t = 1\right]-\E\left[Y_{t+k}\mid A_t = 0, H_t, I_t = 1\right]\mid S_{kt}, I_t=1\big] \\
& &\phantom{bbbbbbbbbbbbbbbbbbbbbbbbb}    -   f_{kt}(S_{kt}) ^\tr\beta_k'\Big).
\end{eqnarray*}
Thus $ \beta_k'=$
\begin{eqnarray}
\label{eq:projbeta}
& &\left[\E\dot U_{\mathrm{W}}(\eta^*,\rho^*)\right]_{(2,2)}^{-1}
  \E\bigg[ \sum_{t=1}^{T-k+1} f_{kt}(S_{kt}) (1-\tilde p_t(1|S_{kt};\rho^*))\tilde p_t(1|S_{kt};\rho^*)I_t \nonumber\\
& &\phantom{bbbbbbbbbbbbbbbbbbbb}\E\Big[\E\left[Y_{t+k}\mid A_t = 1, H_t, I_t = 1\right]-\E\left[Y_{t+k}\mid A_t = 0, H_t, I_t = 1\right]\mid S_{kt}, I_t=1\Big]\bigg].\nonumber\\
& &
\end{eqnarray} 
where 
\begin{eqnarray*}
\left[\E\dot U_{\mathrm{W}}(\eta^*,\rho^*)\right]_{(2,2)}= \E\sum_{t=1}^{T-k+1} f_{kt}(S_{kt}) f_{kt}(S_{kt}) ^\tr (1-\tilde p_t(1|S_{kt};\rho^*))\tilde p_t(1|S_{kt};\rho^*)I_t.
\end{eqnarray*} 

Recall  that  modeling assumption (\ref{eq:withavail}) is,
 $$\E\big[\E\left[Y_{t+k}\mid A_t = 1, H_t, I_t = 1\right]-\E\left[Y_{t+k}\mid A_t = 0, H_t, I_t = 1\right]\mid S_{kt}, I_t=1\big] = f_{kt}(S_{kt}) ^\tr\beta_k^*.$$  
From (\ref{eq:projbeta}), we see that when  modeling assumption (\ref{eq:withavail}) is incorrect then the data analyst's choice of $\tilde p_t(1|S_{kt};\rho^*)$ determines the estimand.  Indeed if the data analyst chooses $\tilde p_t(1|S_{kt};\rho^*)$ to be a constant then,  the limit in probability of $\hat\beta_k$ is given by
\begin{eqnarray}
\label{eq:projbeta1}
 \beta_k'&=&\left[\E\sum_{t=1}^{T-k+1} f_{kt}(S_{kt}) f_{kt}(S_{kt}) ^\tr I_t\right]^{-1}
  \E\bigg[ \sum_{t=1}^{T-k+1} f_{kt}(S_{kt}) I_t \\
& &\phantom{bbbbbbbb}\E\Big[\E\left[Y_{t+k}\mid A_t = 1, H_t, I_t = 1\right]-\E\left[Y_{t+k}\mid A_t = 0, H_t, I_t = 1\right]\mid S_{kt}, I_t=1\Big]\bigg].\nonumber
\end{eqnarray} 
In the case in which $ f_{kt}(S_{kt})=1$ (i.e., $ S_{kt}=\emptyset$) then the scalar  estimand, $\beta_k'$, is simply an average (weighted by availability) of proximal treatment effects:
\begin{eqnarray}
\label{eq:simpleprojbeta}
 \frac{  \sum_{t=1}^{T-k+1}\E[I_t] \E\Big[\E\left[Y_{t+k}\mid A_t = 1, H_t, I_t = 1\right]-\E\left[Y_{t+k}\mid A_t = 0, H_t, I_t = 1\right]\mid  I_t=1\Big]}{ \sum_{t=1}^{T-k+1} \E[I_t]}.
\end{eqnarray}

\section{Additional simulation results}
\label{app:sim}

 This section extends the three simulation experiments considered in \cref{sec:sim} (which focused on $n=T=30$) to different sample sizes $n$ and number of time points $T$. 
Specifically, \cref{tab:sim:omit:supplement}, \cref{tab:sim:stable:supplement}, and \cref{tab:sim:ar1:supplement} below are extensions of \cref{tab:sim:omit}, \cref{tab:sim:stable}, and \cref{tab:sim:ar1}, respectively, for the different combinations of $n=30,60$ with $T=30,50$.   In addition, in order to examine the performance of our estimator of the standard error, we provide the Monte Carlo standard deviation of the point estimates (SD) and the Monte Carlo average standard error estimates (SE) for the weighted and centered estimator for all scenarios considered (the SE statistic was not provided in  \cref{sec:sim}). 

For the first simulation experiment concerning the estimation of a marginal proximal effect when an important moderator exists, see \cref{tab:sim:omit:supplement}: In terms of bias, results were similar to those reported in \cref{sec:sim} for different values of $n$ and $T$. As before, the weighted and centered estimator was unbiased for all values of $\beta^*_{11}$, whereas the bias of the GEE-IND and GEE-AR(1) estimators increased as the magnitude of the underlying effect moderator $\beta^*_{11}$ increased. In terms of $95\%$ confidence intervals, we note that in these simulations coverage probabilites for the GEE-IND and GEE-AR(1) estimators generally worsen for larger values of $n$ and $T$. Finally, in all cases, the average of the standard errors of the proposed weighted and centered estimator closely approximated the Monte Carlo SD.

\begin{table}[H]
\caption{Comparison of three estimators of the marginal proximal treatment effect, $\hat\beta_1$, when an important moderator is omitted.}
\label{tab:sim:omit:supplement}
{\centering
\begin{tabular}{l ccccc cccc cccc}
	\toprule
	& \multicolumn{5}{c}{Weighted and Centered} & \multicolumn{4}{c}{GEE-IND} & \multicolumn{4}{c}{GEE-AR(1)} \\
	\cmidrule(lr){ 2-6 }\cmidrule(lr){ 7-10 }\cmidrule(lr){ 11-14 } 
	 &   &   &   & Root &  &        & & Root &                 & & & Root &  \\ 
	$\beta^*_{11}$ & Mean & SD & SE & MSE & CP &                    Mean & SD    & MSE &        CP             & Mean & SD &   MSE &        CP \\ 
	\midrule
	\multicolumn{14}{c}{ $n=T=30$} \\
	$0.2$          & --0.20 & 0.08 & 0.08 & 0.08 & 0.96  & \bftab --0.17 & 0.07 & 0.07 &        0.94 & \bftab --0.16 & 0.04 & 0.06 & \bftab 0.86 \\ 
	$0.5$          & --0.20 & 0.08 & 0.08 & 0.08 & 0.95  & \bftab --0.14 & 0.07 & 0.09 & \bftab 0.88 & \bftab --0.13 & 0.05 & 0.09 & \bftab 0.70 \\ 
	$0.8$          & --0.20 & 0.08 & 0.08 & 0.08 & 0.95  & \bftab --0.10 & 0.07 & 0.12 & \bftab 0.78 & \bftab --0.10 & 0.05 & 0.12 & \bftab 0.57 \\ 
	\midrule
	\multicolumn{14}{c}{$n=30, T=50$} \\
	$0.2$          & --0.20 & 0.06 & 0.06 & 0.06 & 0.95  & \bftab --0.17 & 0.05 & 0.06 & \bftab 0.92 & \bftab --0.16 & 0.03 & 0.05 & \bftab 0.73 \\ 
	$0.5$          & --0.20 & 0.06 & 0.06 & 0.06 & 0.95  & \bftab --0.14 & 0.06 & 0.08 & \bftab 0.80 & \bftab --0.13 & 0.04 & 0.08 & \bftab 0.49 \\ 
	$0.8$          & --0.20 & 0.07 & 0.07 & 0.07 & 0.94  & \bftab --0.11 & 0.06 & 0.11 & \bftab 0.64 & \bftab --0.10 & 0.04 & 0.11 & \bftab 0.32 \\ 
	\midrule
	\multicolumn{14}{c}{$n=60, T=30$} \\
	$0.2$          & --0.20 & 0.06 & 0.05 & 0.06 & 0.95  & \bftab --0.17 & 0.05 & 0.06 & \bftab 0.90 & \bftab --0.16 & 0.03 & 0.05 & \bftab 0.72 \\ 
	$0.5$          & --0.20 & 0.06 & 0.06 & 0.06 & 0.95  & \bftab --0.14 & 0.05 & 0.08 & \bftab 0.76 & \bftab --0.13 & 0.03 & 0.08 & \bftab 0.41 \\ 
	$0.8$          & --0.20 & 0.06 & 0.06 & 0.06 & 0.94  & \bftab --0.11 & 0.06 & 0.11 & \bftab 0.56 & \bftab --0.10 & 0.04 & 0.11 & \bftab 0.25 \\ 
	\midrule

	\multicolumn{14}{c}{$n=60, T=50$} \\
	$0.2$          & --0.20 & 0.04 & 0.04 & 0.04 & 0.94  & \bftab --0.17 & 0.04 & 0.05 & \bftab 0.87 & \bftab --0.16 & 0.02 & 0.05 & \bftab 0.55 \\ 
	$0.5$          & --0.20 & 0.04 & 0.04 & 0.04 & 0.95  & \bftab --0.14 & 0.04 & 0.07 & \bftab 0.59 & \bftab --0.13 & 0.02 & 0.08 & \bftab 0.19 \\ 
	$0.8$          & --0.20 & 0.05 & 0.05 & 0.05 & 0.95  & \bftab --0.10 & 0.04 & 0.11 & \bftab 0.33 & \bftab --0.10 & 0.03 & 0.11 & \bftab 0.06 \\ 
	\bottomrule
\end{tabular}
\par}\smallskip
\begin{minipage}{0.95\linewidth}
MSE, mean squared error, and SD, standard deviation of $\hat\beta_1$; SE, average of the standard errors for the weighted and centered estimator; CP, $95$\% confidence interval coverage probability for $\beta_1^*=-0.2$.  Results are based on $1000$ replicates. Boldface indicates whether Mean or CP are significantly different, at the $5$\% level, from -0.2 or 0.95, respectively.  
\end{minipage}
\end{table}

For the second simulation experiment concerning the stabilization of the weights $W_t$ in the proposed approach, see \cref{tab:sim:stable:supplement}: results were similar to those reported in \cref{sec:sim}. In all cases, the average of the standard errors of the proposed weighted and centered estimator closely approximated the Monte Carlo SD.

\begin{table}[H]
\caption{Weighted and centered estimator of the marginal proximal treatment effect, $\hat\beta_1$, using two choices for $\tilde{p}_t$.}
\label{tab:sim:stable:supplement}
{\centering
\begin{tabular}{ll ccccc ccccc}
	\toprule
	& & \multicolumn{5}{c}{$\tilde{p}_t$ is constant (i)} & \multicolumn{5}{c}{$\tilde{p}_t$ depends on $S_t$ (ii)} \\
	\cmidrule(lr){ 3-7 }\cmidrule(lr){ 8-12 } 
	$n$ & $T$         & Mean           & SD   & SE & RMSE & CP                & Mean             & SD   & SE   & RMSE & CP \\ 
	\midrule
	30 & 30           &       --0.20   & 0.08 & 0.08 & 0.08 & 0.94            & \bftab  --0.14   & 0.09 & 0.09 & 0.11 & \bftab 0.89 \\ 
	   & 50           &       --0.20   & 0.06 & 0.06 & 0.06 & 0.95            & \bftab  --0.14   & 0.07 & 0.07 & 0.09 & \bftab 0.86 \\
	60 & 30           &       --0.20   & 0.06 & 0.06 & 0.06 & 0.95            & \bftab  --0.14   & 0.06 & 0.06 & 0.09 & \bftab 0.83 \\ 
	   & 50           &       --0.20   & 0.04 & 0.04 & 0.04 & 0.94            & \bftab  --0.14   & 0.05 & 0.05 & 0.08 & \bftab 0.72 \\
	\bottomrule
\end{tabular}\par}
\smallskip
\begin{minipage}{0.95\linewidth}
RMSE, root mean squared error, and SD, standard deviation of $\hat\beta_1$; SE, average of the standard errors for the proposed estimator with appropriate $\tilde{p}_t$; CP, $95$\% confidence interval coverage probability for $\beta_1^*=-0.2$.  Results are based on $1000$ replicates.  Boldface indicates whether Mean or CP are significantly different, at the $5$\% level, from -0.2 or 0.95, respectively.
\end{minipage}
\end{table}

For the third simulation experiment concerning the use of a non-independent working correlation structure in the proposed approach, see \cref{tab:sim:ar1:supplement}: results were similar to those in \cref{sec:sim}, with worsening CP under the non-independent working correlation for larger $n$. In all cases, the average of the standard errors of the proposed weighted and centered estimator (with an independent working correlation) closely approximated the Monte Carlo SD.

\begin{table}[H]
\caption{Weighted and centered estimator of the marginal proximal effect, $\hat\beta_1$, with different working correlation structures. }
\label{tab:sim:ar1:supplement}
{\centering
\begin{tabular}{ll ccccc cccc}
	\toprule
	& & \multicolumn{5}{c}{Independent working correlation (i)} & \multicolumn{4}{c}{AR(1) working correlation     (ii)} \\
	\cmidrule(lr){ 3-7 }\cmidrule(lr){ 8-11 } 
	$n$ & $T$         & Mean           & SD   & SE   & RMSE & CP            & Mean           & SD   & RMSE & CP \\ 
	\midrule
	30 & 30           &       --0.20   & 0.07 & 0.07 & 0.07 & 0.96          & \bftab--0.13   & 0.06 & 0.09 & \bftab 0.66 \\ 
	   & 50           &       --0.20   & 0.05 & 0.05 & 0.05 & 0.96          & \bftab--0.13   & 0.03 & 0.07 & \bftab 0.47 \\
	60 & 30           &       --0.20   & 0.05 & 0.05 & 0.05 & 0.94          & \bftab--0.14   & 0.03 & 0.07 & \bftab 0.42 \\ 
	   & 50           &       --0.20   & 0.04 & 0.04 & 0.04 & 0.95          & \bftab--0.13   & 0.02 & 0.07 & \bftab 0.16 \\
	\bottomrule 
\end{tabular}\par}
\smallskip
\begin{minipage}{0.95\linewidth}
RMSE, root mean squared error, and SD, standard deviation of $\hat\beta_1$; SE, average of the standard errors for the proposed estimator with independent working correlation; CP, $95$\% confidence interval coverage probability for $\beta_1^*=-0.2$.  Results are based on $1000$ replicates.  Boldface indicates whether Mean or CP are significantly different, at the $5$\% level, from -0.2 or 0.95, respectively.
\end{minipage}
\end{table}

\section{Code to Generate Simulation Results}
\label{app:code}

The R code used to generate the simulation experiment results in this paper can be obtained from
\if1\blind
\emph{[Link suppressed to maintain author anonymity]}
\else
 \url{https://github.com/dalmiral/mHealthModeration}.
\fi
This includes the additional calculations necessary to correct standard errors for small samples and for estimated weights (i.e., when either  $\tilde p_t(1 \mid S_{kt})$ or $ p_t(1 \mid H_t)$ is estimated).

\end{document}